\documentclass{jfm}

\usepackage{graphicx}
\usepackage{natbib,color}
\usepackage{float}
\usepackage{amsmath,amssymb,stmaryrd} 
\usepackage{array}
\usepackage{epsfig,color,soul,rotating,overpic}
\usepackage{subfig, setspace}
\usepackage{epstopdf}
\usepackage[section]{placeins}
\usepackage{float}
\usepackage{array}
\usepackage{hyperref}

\ifCUPmtlplainloaded \else
  \checkfont{eurm10}
  \iffontfound
    \IfFileExists{upmath.sty}
      {\typeout{^^JFound AMS Euler Roman fonts on the system,
                   using the 'upmath' package.^^J}%
       \usepackage{upmath}}
      {\typeout{^^JFound AMS Euler Roman fonts on the system, but you
                   dont seem to have the}%
       \typeout{'upmath' package installed. JFM.cls can take advantage
                 of these fonts,^^Jif you use 'upmath' package.^^J}%
      }
  \else
  \fi
\fi

\ifCUPmtlplainloaded \else
  \checkfont{msam10}
  \iffontfound
    \IfFileExists{amssymb.sty}
      {\typeout{^^JFound AMS Symbol fonts on the system, using the
                'amssymb' package.^^J}%
       \usepackage{amssymb}%
       \let\le=\leqslant  
       \let\ge=\geqslant  
      }{}
  \fi
\fi

\ifCUPmtlplainloaded \else
  \IfFileExists{amsbsy.sty}
    {\typeout{^^JFound the 'amsbsy' package on the system, using it.^^J}%
     \usepackage{amsbsy}}
    {\providecommand\boldsymbol[1]{\mbox{\boldmath $##1$}}}
\fi

\newcommand\fig{figure }

\title[Network-theoretic approach to sparsified discrete vortex dynamics]{Network-theoretic approach to sparsified discrete vortex dynamics}

\author[A. G. Nair and K. Taira]
{Aditya G. Nair$^1$
   and 
 Kunihiko Taira$^1$\thanks{Email address for correspondence: ktaira@fsu.edu}}

\affiliation{$^1$Department of Mechanical Engineering and Florida Center for Advanced Aero-Propulsion, Florida State University, Tallahassee, FL 32310, USA\\ [\affilskip]}

\pubyear{2015}
\volume{???}
\pagerange{???--???}
\begin{document}

\maketitle
%=====================================================================
%=====================================================================
%=====================================================================

%All papers should feature a single-paragraph abstract of no more than 250 words, which provides a summary of the main aims and results. 
\begin{abstract}
We examine discrete vortex dynamics in two-dimensional flow through a network-theoretic approach. The interaction of the vortices is represented with a graph, which allows the use of network-theoretic approaches to identify key vortex-to-vortex interactions. We employ sparsification techniques on these graph representations based on spectral theory for constructing sparsified models and evaluating the dynamics of vortices in the sparsified setup. Identification of vortex structures based on graph sparsification and sparse vortex dynamics are illustrated through an example of point-vortex clusters interacting amongst themselves. We also evaluate the performance of sparsification with increasing number of point vortices. The sparsified-dynamics model developed with spectral graph theory requires reduced number of vortex-to-vortex interactions but agrees well with the full nonlinear dynamics. Furthermore, the sparsified model derived from the sparse graphs conserves the invariants of discrete vortex dynamics. We highlight the similarities and differences between the present sparsified-dynamics model and the reduced-order models.
\end{abstract}

%=====================================================================

\begin{keywords}
mathematical foundations, vortex dynamics, vortex interactions
%Authors should not enter keywords on the manuscript, as these must be chosen by the author during the online submission process and will then be added during the typesetting process (see http://journals.cambridge.org/data/\linebreak[3]relatedlink/jfm-\linebreak[3]keywords.pdf for the full list)
\end{keywords}
%\tableofcontents

%=====================================================================
%%%%%%%%%%%%%%%%%%%%%%%%%%%%%%%%%%%%%%%%%%%%%%%%%%%%%%%%

\section{Introduction}
\label{sec:intro}

Network describes how components are linked to one another. We can represent the components by points (nodes) and the connections by lines (edges) through a mathematical abstraction. The structure comprised of these nodes and edges is called a graph, which has been studied in detail in the field of graph theory \citep{Bollobas98}.  Network analysis is concerned with the study of graphs as well as the interaction and evolution of the variables of interest on graphs over time and space \citep{Newman10}.  The framework developed in network analysis and graph theory can provide insights into how the structure of a network can influence the overall dynamics taking place on the network.  

Network analysis and graph theory by nature are very fundamental and generic, which enable them to impact a wide range of applications, including the analysis of biological and social networks, study of traffic flows, and design of robust power grids \citep{Newman10}. Biologists and medical scientists use network analysis to determine how electrical signals travel inside the brain and how abnormality in the brain network connections can affect the normal functionality \citep{DuarteCavajalino:Neuroimage12, Owen:Neuroimage13}. 

In epidemiology, researchers model the outbreak of diseases on the population network. Locations with high concentration of population, such as airports, stations, schools, and hospitals, can be represented on a network with large number of connections. Identifying such locations is especially critical when containment measures are designed to control outbreaks of HIV \citep{Morris:SMR93}, SARS \citep{LloydSmith:Nature05}, and Influenza \citep{Glass:EID06, Cauchemeza:PNAS11}.  Each of these diseases have different dynamics and an associated network structure. Once the high-risk groups and areas are identified, network analysis can assist in designing and implementing prevention and combat strategies in the most swift manner with limited resources \citep{Salathe:PLOSCB10, Robinson:TPB12}. Network analysis can also reveal how a group of people are socially connected to one another and examine how subgroups within a population are interlinked in a complex manner \citep{Porter:PNAS05}. Moreover, network analysis has been utilized in electrical engineering to determine the voltages and currents associated with electrical circuits via graph representations \citep{wai2004electrical}. In the aforementioned applications of network analysis, the connections between people or elements are highlighted. 

In the field of fluid mechanics, there have been extensive studies performed to capture the behavior of complex fluid flow. Lagrangian based methods, such the vortex methods, allow us to simulate the unsteady fluid flow \citep{leonard1980vortex,Cottet00}. These methods involve modeling of fluid flow with point vortices, vortex sheets, vortex filaments or vortex patches \citep{Saffman92,Cottet00}. Low-order representation of these vortex models \citep{wang2013low,hemati2014improving} have been proposed in recent years. The evaluation of the velocity field for vortex methods rely often on fast summation methods \citep{Greengard:JCP87} for reduced computational time. The objective of the present study is to extend network analysis and graph theory to discrete point-vortex dynamics.

As the computational approaches for vortex dynamics are being developed, there are also ongoing efforts in flow modeling. Reduced-order models have been utilized successfully to describe unsteady incompressible and compressible flows \citep{Rowley2004model,Noack:JFM05}. One such approach is to utilize Galerkin projection to derive reduced-order models using spatial bases such as the Proper Orthogonal Decomposition (POD) modes \citep{berkooz1993proper,Holmes96}. The reduced-order model distills the infinite-dimensional Navier-Stokes equations to model equations with state variables having significantly reduced dimensions. In the present work, we also aim to capture the essential physics of unsteady fluid flow but not by reducing the dimension of the state variable. Instead, we examine the interaction between the elements of the state variables and {\it sparsify the interactions} utilizing a network-theoretic approach. We refer to the dynamical model based on sparsifying the interactions as {\it sparsified dynamics} in this paper.

In the present work, we consider representing vortices with nodes and the interactions amongst the vortices with edges.  By utilizing the network-theoretic framework to study vortex dynamics, we highlight the connections (edges) that the vortices have in the flow field.  Such analysis emphasizes how a collection of vortices influence each other through a causal point of view on a network structure. We believe that the present study can provide an alternative tool to analyze how vortices or flow structures interact in the flow field and support the development of interaction-based models to capture unsteady vortex dynamics. Furthermore, we consider the use of graph sparsification as a tool for sparsifying the interaction between the point vortices. These models keep the nodes intact and reduce the number of edges maintaining the dimensionality of the original system unlike reduced-order models. The removal of edges can drastically reduce computational cost to model the full dynamical behavior, sharing the same spirit as reduced-order models. 

In what follows, we first introduce network analysis and graph theory briefly in \S{\ref{sec:ntf}}.  We introduce the notion of graph sparsification in \S{\ref{sec:gs}}. We then consider the use of network analysis on discrete vortex dynamics and derive sparsified models in \S{\ref{sec:advd}}. Some comments are offered on the difference between the sparsified-dynamics model and the reduced-order model.  Concluding remarks are offered in \S{\ref{sec:conc}}.  

%%%%%%%%%%%%%%%%%%%%%%%%%%%%%%%%%%%%%%%%%%%%%%%%%%%%
%\vspace{-3 mm}
\section{Network-theoretic framework}
\label{sec:ntf}

\begin{figure}
   \begin{center}
     \begin{tabular}{ccc}
       Example graph & Complete graph $K_5$\\
        \begin{overpic}[height=0.35\textwidth]{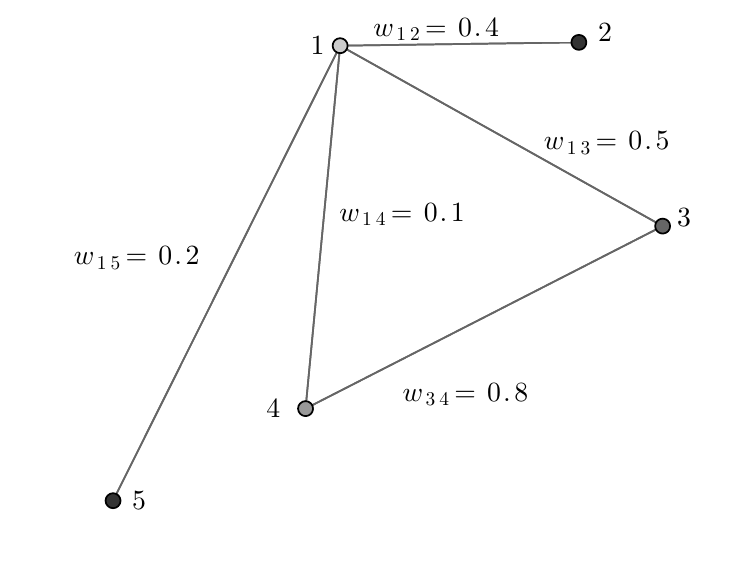}
        \end{overpic} &
        \begin{overpic}[height=0.35\textwidth]{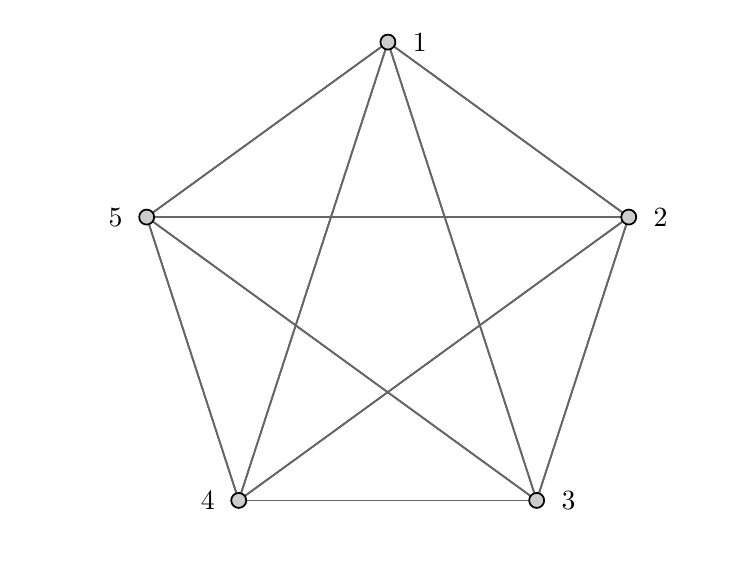}
        \end{overpic} 
   \end{tabular}
   \end{center}
   \caption{(left) An example weighted graph; and (right) a complete graph $\mathcal{K}_5$.}
   \label{fig_nod}
\end{figure}

A network is defined by a collection of vertices joined by edges. These edges form the bonds that connect the vertices or the various entities of the system together. The edges can be associated with weights describing the importance of the connections between the vertices. Such a graph is called as a weighted graph. If the edges of the graph have an associated directivity (i.e., one vertex influences the other but not vice versa), the graph is called a directed graph. If a vertex influences itself, a self-loop edge (edge reconnected to itself) can represent such an effect. Any undirected graph $\mathcal{G}$ can be described by a set of vertices $V=\{v_1,v_2,...v_N\}$, a set of edges $E$, and a set of weights $w$ associated with the edges, i.e., $\mathcal{G} = \{V,E,w\}$ \citep{Chung:1997, Newman10}. An example of a weighted graph is shown in \fig \ref{fig_nod} (left) with weights associated with the edges displayed. A complete graph $\mathcal{K}_N$ with $N = 5$, shown in \fig \ref{fig_nod} (right), has all the vertices connected to each other. In other words, this graph has $N$ vertices with complete set of possible $N(N-1)/2$ edges without any self-loops. We later use a weighted version of the complete graph to describe the interactions amongst a set of discrete point vortices in the context of fluid dynamics.

Network connections can be summarized by its adjacency matrix $\boldsymbol{A}_{\mathcal{G}} \in \mathbb{R}^{N\times N}$, which is given by
\begin{equation}
[\boldsymbol{A}_{\mathcal{G}}]_{ij} = 
\begin{cases}
	w_{ij} &\text{if } (i,j) \in E \\
         0 &\text{otherwise.}
\end{cases}
\label{adjeq}
\end{equation}
In an unweighted graph, the weight is set to unity if $(i,j)\in E$. The diagonal entries of the adjacency matrix relate to the weight of the self-connecting (loops) edges. The adjacency matrix of an undirected graph is symmetric. The degree $k_i$ of a vertex $i$ represents the summation of the weights of the edges connected to it given by $k_i = \sum_{j=1}^N [{\boldsymbol{A}}_\mathcal{G}]_{ij}$. The adjacency matrix for an example graph ${\boldsymbol{A}}_{\mathcal{G}}$ and the complete graph ${\boldsymbol{A}}_{{\mathcal{K}}_5}$ shown in \fig \ref{fig_nod} are given by
\[
\boldsymbol{A}_{\mathcal{G}} = \left( \begin{array} {ccccc}
0 & 0.4 & 0.5 & 0.1 & 0.2 \\
0.4 & 0 & 0 & 0 & 0 \\
0.5 & 0 & 0 & 0.8 & 0 \\
0.1 & 0 & 0.8 & 0 & 0 \\
0.2 & 0 & 0 & 0 & 0 
\end{array} \right) \text{and } 
{\boldsymbol{A}}_{{\mathcal{K}_5}} = \left( \begin{array} {ccccc}
0 & 1 & 1 & 1 & 1 \\
1 & 0 & 1 & 1 & 1 \\
1 & 1 & 0 & 1 & 1 \\
1 & 1 & 1 & 0 & 1 \\
1 & 1 & 1 & 1 & 0 
\end{array} \right),
\]
where $\mathcal{G}$ is a weighted graph and $\mathcal{K}_5$ is an unweighted complete graph. 

Another important matrix in graph theory is the graph Laplacian matrix $\boldsymbol{L}_{\mathcal{G}} \in \mathbb{R}^{N\times N}$, which is given by
\begin{equation}
[{\boldsymbol{L}}_{{\mathcal{G}}}]_{ij} =
\begin{cases}
	 k_i &\text{if } (i,j) \in E \text{ and }  (i = j) \\
         -w_{ij}  &\text{if } (i,j) \in E \text{ and }  (i \neq j) \\
         0  &\text{otherwise.}
\end{cases}
\label{lapeq}
\end{equation}
The graph Laplacian matrix can also be deduced from the adjacency matrix, $\boldsymbol{L}_{{\mathcal{G}}} = \boldsymbol{D}_{{\mathcal{G}}} - \boldsymbol{A}_{\mathcal{G}}$, where $\boldsymbol{D}_{{\mathcal{G}}} \in \mathbb{R}^{N\times N}$ is a diagonal matrix with elements equal to degrees of vertices, $\boldsymbol{D}_{{\mathcal{G}}} = \text{diag}([k_i]_{i=1}^N)$. For undirected networks, it is symmetric and positive semidefinite (singular). This Laplacian matrix is a discrete analog of the negative continuous Laplacian operator ($-\nabla^2$) and is also called as the discrete Laplacian. It is naturally defined by its quadratic form. If we have a linear system of the form $\boldsymbol{L}_{{\mathcal{G}}} \boldsymbol{x} = \boldsymbol{b}$, for the vector $\boldsymbol{x} \in \mathbb{R}^N$, the Laplacian quadratic form of a weighted graph $\mathcal{G}$ is given by \citep{mohar1991laplacian}
\begin{equation}
\boldsymbol{x}^T{\boldsymbol{L}}_{{\mathcal{G}}}\boldsymbol{x} = \sum_{(i,j)\in E} w_{ij} (x_i - x_j)^2 .
 \label{eqlpp}
\end{equation}  
The discrete Laplacian is a smoothness indicator of $\boldsymbol{x}$ over the edges in $\mathcal{G}$. The Laplacian quadratic form becomes large as $\boldsymbol{x}$ jumps over the edges of $\mathcal{G}$. The definitions of adjacency matrix and graph Laplacian form the building blocks of the graph-theoretic framework. The Laplacian quadratic form will be used below to introduce the notion of spectral similarity of graphs and graph sparsification.

%%%%%%%%%%%%%%%%%%%%%%%%%%%%%%%%%%%%%%%%%%%%%%%%%%%%%%%%
 
\section{Graph sparsification}
\label{sec:gs}

%%%%%%%%%%%%%%%%%%%%%%%%%%%%%%%%%%%%%%%%%%%%%%%%%%%%%%%%

The sparse approximation of dense graphs can provide numerous benefits for numerical computation. Graph sparsification is useful for designing computationally efficient algorithms and helps in identifying representative edges and associated weights \citep{Spielman:SIAMJC11a}. Graph similarity, which forms the basis for graph sparsification, can be derived in a number of ways depending on the desired similarity properties. Distance similarity of graphs \citep{peleg1989optimal} can be achieved by sparse graphs called spanners that keep the same shortest-path distance between each pair of vertices as in the original graph while cut similarity \citep{Benczur96} can be achieved by maintaining the weight of the edges leaving a set of vertices of the sparse graph to be the same as that of the original graph. A much stronger notion of similarity referred to as {\it spectral similarity} was introduced by \cite{Spielman:SIAMJC11a}. Spectral similarity is closely tied to the Laplacian quadratic form of the graphs as defined by equation (\ref{eqlpp}). The concept of spectral similarity directly leads to spectral sparsification of graphs; that is to create sparse graphs, which are spectrally similar to the original graph. 

%%%%%%%%%%%%%%%%%%%%%%%%%%%%%%%%%%%%%%%%%%%%%%%%%%%%%%%%

\subsection{Spectral sparsification}
\label{sec:esparse}

Spectral sparsification is a more general abstraction than cut sparsifiers and maintains spectral similarity between the sparsified and original graphs. In particular, spectral sparsification can remove some of the edges in the graph, while maintaining similar adjacency and Laplacian eigenspectra. One of the key features of spectral sparsification is that it keeps the sum of the weights leaving the vertex of a graph constant. A spectral sparsifier is a subgraph of the original graph whose Laplacian quadratic form is approximately the same as the original graph \citep{Spielman:SIAMJC11a}. 

Sparsification involves the creation of a sparse graph $\mathcal{G}_S$ from the original graph $\mathcal{G}$  based on an approximation order of $\epsilon$. The quadratic form induced by graph Laplacian of $\mathcal{G}$ is maintained upto a multiplicative $(1\pm\epsilon)$ factor by spectral sparsification \citep{kelner2011spectral}. Thus, the sparse graph $\mathcal{G}_S$ is a $(1\pm\epsilon)$-spectral approximation of $\mathcal{G}$. The approximation order $\epsilon$ can vary from zero to unity. The approximation with $\epsilon = 0$ indicates that $\mathcal{G}_S$ is same as original graph $\mathcal{G}$ and none of the edges are sparsified, while $\epsilon = 1$ relaxes the quadratic form induced by the sparse graph to within twice of that induced by the original graph. An approximation of $\epsilon = 1$ leads to a heavily sparsified graph. Denoting the Laplacian matrices of $\mathcal{G}$ and $\mathcal{G}_S$ by ${\boldsymbol{L}}_\mathcal{G}$ and $\boldsymbol{L}_{{\mathcal{G}_S}}$, respectively, the spectrally sparsified Laplacian satisfies
\begin{equation}
(1-\epsilon)\boldsymbol{x}^T\boldsymbol{L}_{{\mathcal{G}}}\boldsymbol{x} 
\le \boldsymbol{x}^T\boldsymbol{L}_{{\mathcal{G}_S}}\boldsymbol{x} 
\le (1+\epsilon)\boldsymbol{x}^T\boldsymbol{L}_{{\mathcal{G}}}\boldsymbol{x}
 \label{lapquad}
\end{equation}
at least with probability 1/2 with large $N$ for all $\boldsymbol{x}\in \mathbb{R}^N$ \citep{Spielman:SIAMJC11b}. For the example problem considered later, these bounds are much tighter. This tells us that $\boldsymbol{L}_{{\mathcal{G}_S}}$ holds eigenvalues similar to those of $\boldsymbol{L}_{{\mathcal{G}}}$. These spectrally similar sparse graphs are found using the spectral sparsification algorithm based on sampling by effective resistance discussed below.

Before we discuss how a graph can be sparsified, let us first follow the works of \cite{Bollobas98} and \cite{srivastava2010spectral} to establish an analogy of a graph to an electrical circuit. If the entire graph is viewed as a resistive circuit, we can define a resistance on the individual edges $e=(i,j)$ of the graph. According to Thomson's principle, the potentials and currents in a resistive circuit distribute themselves so as to minimize the total energy in the network \citep{Bollobas98}. This energy minimization principle leads to the concept of effective resistance.

Effective resistance between vertices $i$ and $j$ is the potential difference induced between them when a unit current is injected at one vertex and extracted at the other \citep{Bollobas98, srivastava2010spectral}. The effective graph resistance (also called as resistance distance) is the sum of the effective resistance over all the pairs of vertices in the graph $\mathcal{G}$ \citep{klein1993resistance,ellens2011effective}. Rayleigh's monotonicity law states that pairwise effective resistance is a non-increasing function of the edge weights \citep{van2011graph}. 

In order to obtain an expression of effective resistance for graph sparsification, we orient the edges of the original weighted undirected graph $\mathcal{G}$ with $N$ vertices and $M$ edges. We can represent any directed graph by a signed edge ($e$)-vertex ($v$) incidence matrix $\boldsymbol{B}_{{\mathcal{G}}} \in \mathbb{R}^{M\times N}$ given by  
\begin{equation}
[\boldsymbol{B}_{{\mathcal{G}}}]_{ev} =
\begin{cases}
	 1 &\text{if } e \in E \text{ and } v \text{ is the head of } e\\
         -1  &\text{if } e \in E \text{ and } v \text{ is the tail of } e\\
         0  &\text{otherwise.}
\end{cases}
\end{equation}
The row of $\boldsymbol{B}_{{\mathcal{G}}}$ corresponding to an edge $e = (i,j)$ is given by $(\boldsymbol{p}_i-\boldsymbol{q}_j)$, where $\boldsymbol{p}_i$ and $\boldsymbol{q}_j$ are elementary unit vectors in the $i$ and $j$ directions, respectively. If the edge weights of the graph are represented in a diagonal matrix given by $\boldsymbol{C}_{{\mathcal{G}}} \in \mathbb{R}^{M\times M}$, we can express the Laplacian matrix based on the incidence matrix as
\begin{equation}
   \boldsymbol{L}_{{\mathcal{G}}} =
   \boldsymbol{B}_{{\mathcal{G}}}^T 
   \boldsymbol{C}_{{\mathcal{G}}} 
   \boldsymbol{B}_{{\mathcal{G}}} = 
   \sum_{i,j \in E}w_{ij}(\boldsymbol{p}_i-\boldsymbol{q}_j)(\boldsymbol{p}_i-\boldsymbol{q}_j)^T.
\label{lap_alt_def}
\end{equation}\\
The above relation holds true only for undirected graphs. The incidence matrix $\boldsymbol{B}_{{\mathcal{G}}}$ and the diagonal matrix $\boldsymbol{C}_{{\mathcal{G}}}$ for the example graph shown in \fig \ref{fig_nod} (left) are given by
\[
\boldsymbol{B}_{\mathcal{G}} = \left( \begin{array} {ccccc}
1 & -1 & 0 & 0 & 0 \\
1 & 0 & -1 & 0 & 0 \\
1 & 0 & 0 & -1 & 0 \\
0 & 0 & 1 & -1 & 0 \\
1 & 0 & 0 & 0 & -1
\end{array} \right) \text{and } 
\boldsymbol{C}_{{\mathcal{G}}}  = \left( \begin{array} {ccccc}
0.4 & 0 & 0 & 0 & 0 \\
0 & 0.5 & 0 & 0& 0 \\
0 & 0 & 0.1 & 0 & 0 \\
0 & 0 & 0 & 0.8 & 0 \\
0 & 0 & 0 & 0 & 0 .2
\end{array} \right).
\]
For undirected graphs, each edge is counted only once in the incidence matrix and the convention for head and tail of an edge can be fixed arbitrarily. Here, for edge $e = (1,2)$ of the example graph shown in \fig \ref{fig_nod} (left), the head and tail are considered to be vertex $1$ and $2$, respectively. The corresponding elementary unit vectors for the edge are given by $p_1 = (1,0,0,0,0)$ and $q_2 = (0,1,0,0,0)$, which is apparent from the first row of $\boldsymbol{B}_{{\mathcal{G}}}$ being $\boldsymbol{p}_1-\boldsymbol{q}_2$.

For an edge corresponding to $e = (i,j)$, the unit current is injected at vertex $i$ and extracted at vertex $j$. Thus, we set the electrical current across the edge to be $(\boldsymbol{p}_i-\boldsymbol{q}_j)$. The potential induced by this current at the vertices is given by $\boldsymbol{L}_{{\mathcal{G}}}^+(\boldsymbol{p}_i-\boldsymbol{q}_j)$, where $\boldsymbol{L}_{{\mathcal{G}}}^+$ is the Moore--Penrose pseudoinverse of  the Laplacian matrix $\boldsymbol{L}_{{\mathcal{G}}}$ \citep{srivastava2010spectral}. The potential difference across edge $e = (i,j)$ is then given by $(\boldsymbol{p}_i-\boldsymbol{q}_j)^T\boldsymbol{L}_{{\mathcal{G}}}^+(\boldsymbol{p}_i-\boldsymbol{q}_j)$. Thus, for unit current, the effective resistance across edge $e=(i,j)$ which corresponds to the potential difference can be expressed as 
\begin{align}
 & [{R}_e]_{ij}= (\boldsymbol{p}_i-\boldsymbol{q}_j)^TL_\mathcal{G}^+(\boldsymbol{p}_i-\boldsymbol{q}_j). 
 \label{eq5}
\end{align} \\ 
The above expression is used for computing effective resistance of the edges of the graph. The sparsification of the original graph $\mathcal{G} = \{V,E,w\}$ is performed with an algorithm $\tt{Sparsify}$ \citep{Spielman:SIAMJC11b} to produce a sparse graph $\mathcal{G}_S = \{V,\widetilde{E},\widetilde{w}\}$ where $\widetilde{w}$ are the weights corresponding to the sparse graph $\mathcal{G}_S$. This algorithm is based on the concept of effective resistance and yields a $(1+\epsilon)$ sparse graph $\mathcal{G}_S$. This sparse graph contains a reduced number of $\mathcal{O}(N\log (N)/\epsilon^2)$ edges. The procedure for sparsification $\mathcal{G}_S = \tt{Sparsify}(\mathcal{G})$ is summarized below.
 
 \subsection{Summary of algorithm}
 
We first create a list of the edges $E$ with the associated weights of the original graph $\mathcal{G}$. The adjacency and Laplacian matrices of graph $\mathcal{G}$ are constructed from equations (\ref{adjeq}) and (\ref{lapeq}). The Moore--Penrose pseudoinverse of the Laplacian matrix $\boldsymbol{L}_{{\mathcal{G}}}^+$ is then computed. For each edge in the edge list, the elementary unit vectors, $\boldsymbol{p}_i$ and $\boldsymbol{q}_j$, and edge weights $w_e$ are obtained. Next, the effective resistance $[{R}_e]_{ij}$ corresponding to each edge is computed from equation (\ref{eq5}). 

A random edge $e=(i,j)$ from the edge list of graph $\mathcal{G}$ is chosen with probability $p_e$ proportional to $w_e [{R}_e]_{ij}$. The edge $e$ is added to the sparse graph $\mathcal{G}_S$ with the weight given by $\widetilde{w}_e = w_e/qp_e$, where $q = 8N\log_2(N)/\epsilon^2$.  We take integer($q$) number of samples independently without replacement and sum the weights if an edge is chosen more than once. The resulting graph becomes the sparsified graph $\mathcal{G}_S$ that satisfies equation ($\ref{lapquad}$) with eigenvalues similar to those of $\mathcal{G}$.

After the random sampling procedure, a sparsified adjacency matrix $\boldsymbol{A}_{{\mathcal{G}_S}}$ is obtained.  For convenience, we define the ratio of the sparsified and original adjacency matrix weights as $W_{ij}$,
\begin{equation}
 W_{ij} \equiv 
 	\begin{cases} 
	{\widetilde{w}_{ij}}/{w_{ij}} & \text{if  $(i,j)\in \widetilde{E}$} \\
	0 & \text{otherwise.} 
	\end{cases}
 \label{eq6}
\end{equation}
This sparsification factor $W_{ij}$ is related to the probability of an edge $e=(i,j)$ of graph $\mathcal{G}$ being sampled. The edges of the original graph that are not sampled during the random sampling procedure have zero weights in the sparsified graph. These edges are cut during sparsification. The spectral sparsification procedure also redistributes the weights of the cut edges among the other edges of the sparsified graph. Thus, cutting of graph edges is compensated by redistribution of the weights to preserve spectral properties of the original graph. 

The spectral sparsification algorithm described above produces a ($1\pm \epsilon$) expander graph, i.e., sparsifier with strong connectivity properties when $\epsilon \ge 1/\sqrt{N}$, where $N$ is the number of vertices \citep{Spielman:SIAMJC11b}. With the concept of graphs and spectral sparsification now discussed, we consider the application of network analysis to discrete vortex dynamics in the next section.

%%%%%%%%%%%%%%%%%%%%%%%%%%%%%%%%%%%%%%%%%

\section{Applications to discrete vortex dynamics}
\label{sec:advd}

We apply network analysis to study the dynamics of a collection of point vortices \citep{Saffman92, Newton01}.  In the present work, the flow is assumed to be two-dimensional, incompressible, and inviscid.  The spatial domain $\mathcal{D}$ is taken to be infinitely large and the position of the vortices in the flow field is denoted by $\boldsymbol{r} = (x,y) \in \mathcal{D}$.  We consider a collection of $N$ vortices to be in the domain $\mathcal{D}$ resulting in the vorticity field of
\begin{equation}
   \omega(\boldsymbol{r}) 
   = \sum_{j=1}^N \frac{\kappa_{j}}{2\pi} \delta(\boldsymbol{r}-\boldsymbol{r}_{j}), \label{eq43}
\end{equation}
where $\kappa_j$ and $\boldsymbol{r}_j$ are the strength (circulation) and position of the $j$-th vortex, respectively, and $\delta(\cdot)$ is the Dirac delta function. The motion of the vortices is described by the Biot--Savart law
\begin{equation}
   \frac{d \boldsymbol{r_i}}{d t} = 
   \sum_{\substack{j=1\\j\ne i}}^N \frac{\kappa_j}{2\pi} \frac{\hat{\boldsymbol{k}} \times 
   (\boldsymbol{r}_i - \boldsymbol{r}_j)}{|\boldsymbol{r}_i - \boldsymbol{r}_j|^2}, \label{eq53}
\end{equation}  
where $\hat{\boldsymbol{k}}$ is the out-of-plane unit normal vector.  To determine the trajectories of the vortices, we numerically integrate the above equation.  This equation is the basis of the discrete vortex methods that are used to simulate unsteady vortical flows in place of discretizing the two-dimensional Euler equations.
    
The objective of the present work is to describe the interactions amongst a set of point vortices with graph theory and network analysis. The dynamics of the vortices will be modeled on a sparsified graph in place of the full (complete) graph to derive a sparsified model that accurately captures the motion of the vortices while conserving physical properties, such as circulation, Hamiltonian, as well as linear and angular impulse.

\subsection{Representing vortex interactions on a graph}

A vortex is advected by the induced velocity from the other vortices and not by its own velocity as seen in equation (\ref{eq53}).  Each vortex is influenced by all other vortices, which means that each and every vortex has a connection to all other vortices without any self-loops.  Thus, the discrete system of $N$ point vortices can be represented by a weighted complete graph, $\mathcal{K}_N$.  The vertices or nodes of the graph represent the discrete point vortices and the edge weights represent the strength of the connections between them.  

The motion of point vortices is influenced by the strengths (circulation) of the individual point vortices and the distances between them. The shorter the distance between the point vortices, the higher is the influence of the vortices on each other. Also, a vortex with higher strength has more ability to influence the surrounding vortices compared to that with lower strength. We assign the weight for the interaction between two vortices to be dependent on the characteristic vortex induced velocity based on the strengths of the two vortices and the distance between them. Thus, this implies that the edge weight for the vortex network should be proportional to $\kappa/r$, where $\kappa$ is the characteristic strength of the two vortices and $r$ is the distance between them within the network.

The distance between the vortices is dependent on the location of the two point vortices $|\boldsymbol{r}_i - \boldsymbol{r}_j|$ and the characteristic strength can be established by considering their geometric mean $\sqrt{|\kappa_i \kappa_j|}$. This choice makes the adjacency matrix symmetric which allows the use of the spectral sparsification algorithm discussed in \S{\ref{sec:esparse}}. Therefore, the adjacency matrix for the graph representation of a set of point vortices can be defined as,
\begin{equation}
 	[{\boldsymbol{A}}_{{\mathcal{G}}}]_{ij} = 
	\begin{cases}
  	 \frac{\sqrt{|\kappa_i\kappa_j|}}{|\boldsymbol{r}_i - \boldsymbol{r}_j|} &\text{if $(i,j) \in E$ and $i \neq j$}\\
         0 &\text{otherwise.}
	\end{cases}
	\label{eq13}
\end{equation}
Note that the diagonal elements of the adjacency matrix are zero as the self-induced velocity components are zero.  This translates to the graph not having any self-loops. The interaction between the system of discrete point vortices represented on a complete graph can be sparsified for deriving the sparse vortex interaction model.

In addition to considering the geometric mean of the circulation for the weights, algebraic mean of the circulation was also considered. The computation using the algebraic mean (i.e., $\frac{1}{2} (|\kappa_i|+|\kappa_j|)/|\boldsymbol{r}_i - \boldsymbol{r}_j|$) for the weights perform similar to that with the above geometric mean.  While we show that geometric mean sufficiently captures the vortex interaction, there may be some room for optimizing the choice of weights. In what follows we utilize the geometric mean in equation (\ref{eq13}) for the network weight.

%%%%%%%%%%%%%%%%%%%%%%%%%%%%%%%%%%%%%%%%%%

\subsection{Sparsification of vortex interactions}
\label{sec:rom}
In this section, we construct sparsified representation of vortex interactions by sparsifying and redistributing the weights of the connections between the point vortices as discussed in \S{\ref{sec:esparse}}. We consider a configuration of $N$ point vortices in $n_c$ clusters. This setup allows us to examine the effect of sparsification on a cluster of vortices. The edge weights represented in the adjacency matrix are given by equation (\ref{eq13}). 

\begin{figure}
   \begin{center}
     \begin{tabular}{c}
       %Vortex distribution\\
        \begin{overpic}[height=0.45\textwidth]{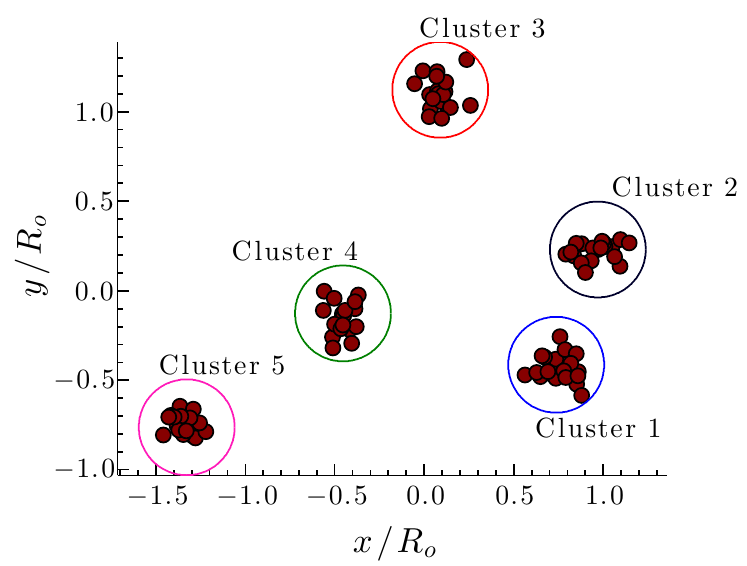}
        \end{overpic} 
   \end{tabular}
   \end{center}
   \caption{The spatial arrangement of $N = 100$ point vortices with five clusters.}
   \label{fig_cluster}
\end{figure}

Let us first consider a collection of $N = 100$ point vortices with $n_c = 5$ with each cluster containing $20$ vortices. The clusters in this setup can be clearly identified by the use of graph clustering algorithms.  An algorithm maximizing modularity \citep{newman2004fast} can be utilized for the identification of clusters. One could also use $k$-means for cluster identification similar to the work performed by \cite{kaiser2013cluster}. The point vortices in a particular cluster are given a normal distribution in space with a radial standard deviation of $\sigma_r = 0.2$. The strength of the point vortices, $\kappa_i$, within the individual clusters have a normal distribution with the mean of $\bar{\kappa} = 0.1$ and a standard deviation of $\sigma_\kappa = 0.01$. The setup for this configuration is shown in figure \ref{fig_cluster}. The positions of the clusters are normalized by the average radial distance of the centroid of the clusters from the geometric center of the system ($R_o$).

\begin{figure}
   \begin{center}
     \begin{tabular}{c|cc}
      & \bf{Vortex Network} & \bf{Sparsity pattern} \\ \hline 
        \begin{sideways} \hspace{0.12\textwidth} Original Network \end{sideways} &
        \includegraphics[width=0.42\textwidth]{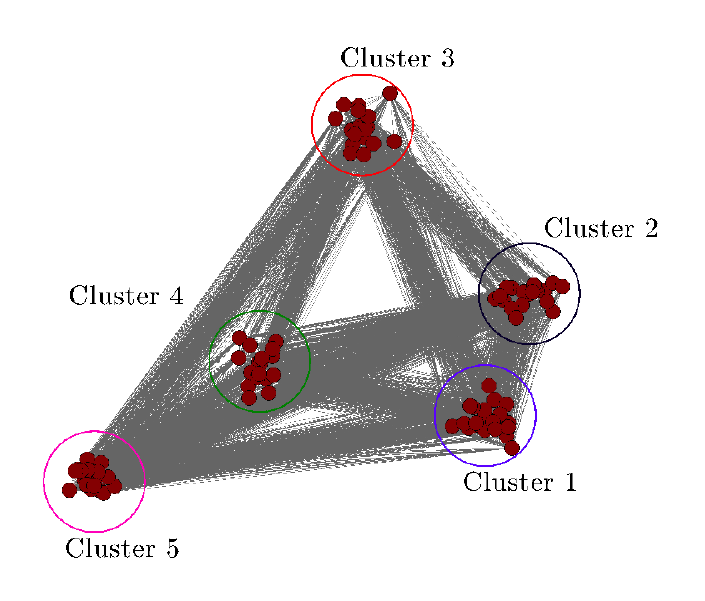} &
        \includegraphics[width=0.42\textwidth]{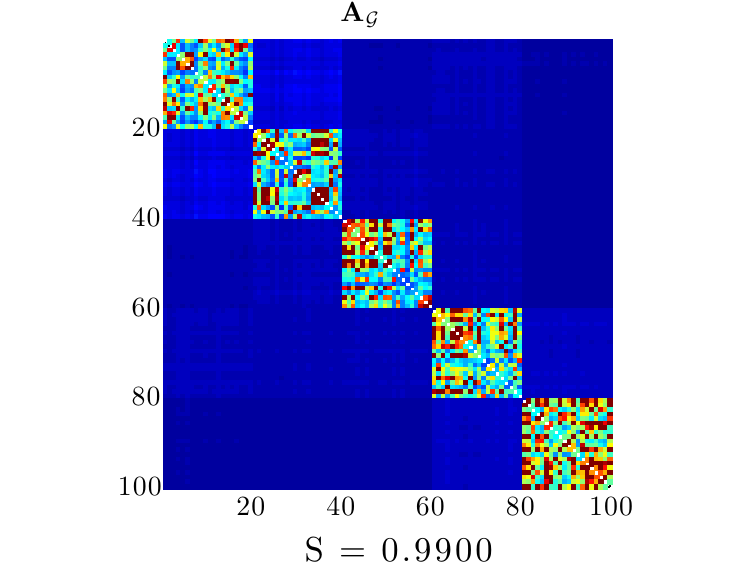} \\ 
        \begin{sideways} \hspace{0.05\textwidth} Sparsified Network ($\epsilon = 0.5$) \end{sideways} &
        \includegraphics[width=0.42\textwidth]{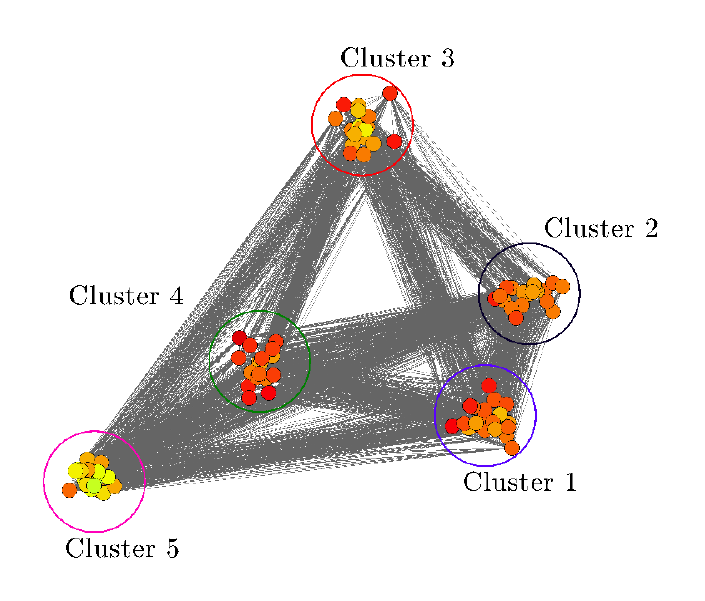} &
        \includegraphics[width=0.42\textwidth]{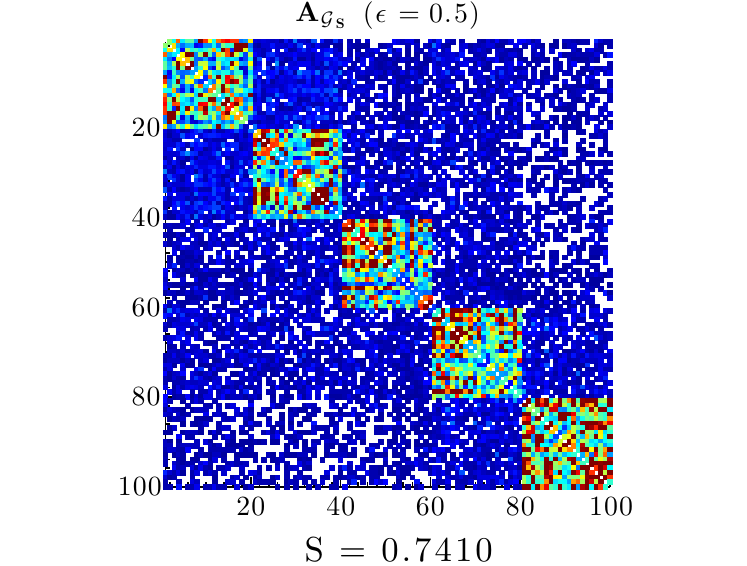} \\
        \begin{sideways} \hspace{0.05\textwidth} Sparsified Network ($\epsilon = 1$) \end{sideways} &
        \includegraphics[width=0.42\textwidth]{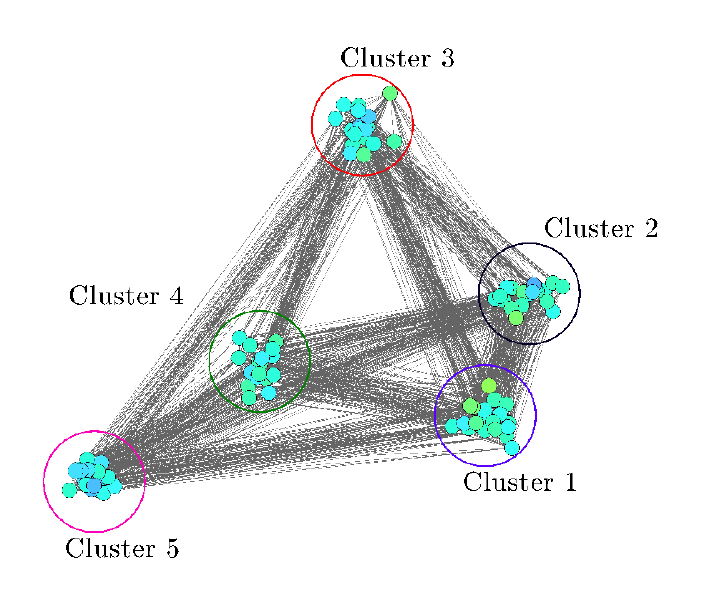} &
        \includegraphics[width=0.42\textwidth]{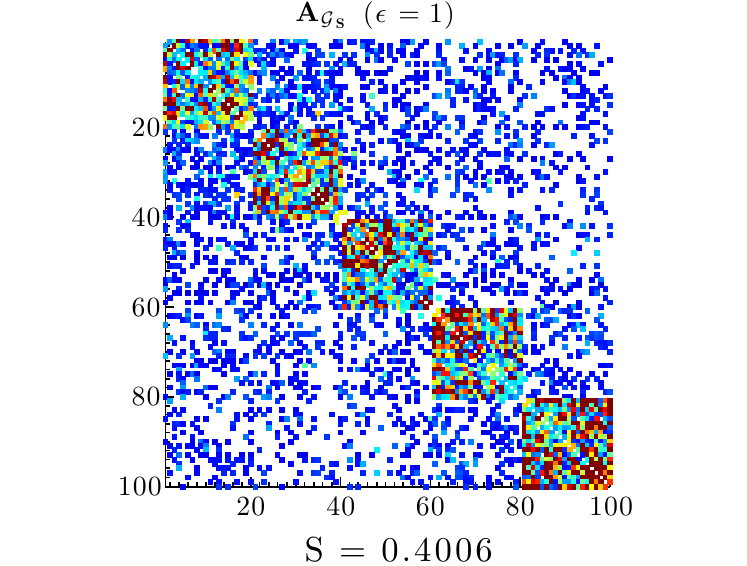} \\
        \begin{sideways} \hspace{0.12\textwidth}  \end{sideways} &
        \includegraphics[width=0.48\textwidth]{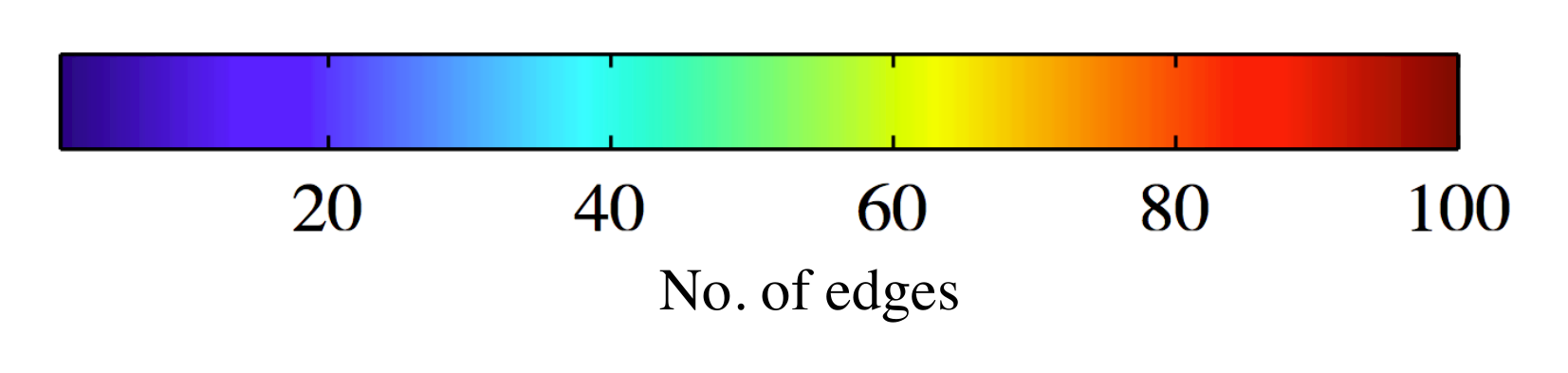} &
        \includegraphics[width=0.22\textwidth]{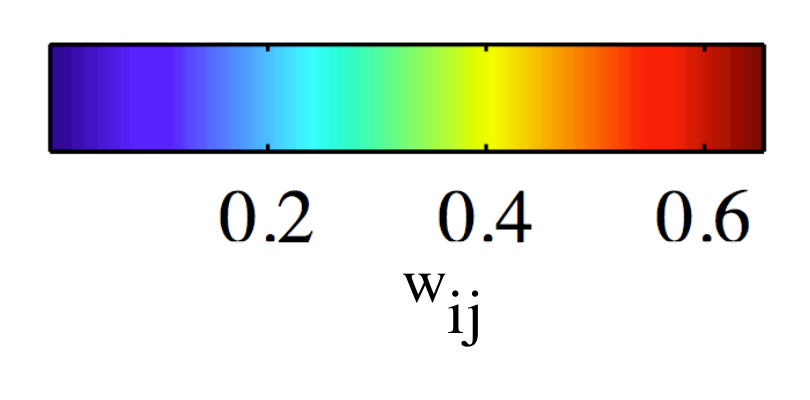} 
        \end{tabular}
   \end{center}
   \caption{Original and sparsified graphs and their respective sparsity patterns of the adjacency matrix for $N = 100$ vortices. The color of the nodes of the vortex network represents the unweighted connections (number of connected edges) of the individual point vortices. The color in the sparsity patterns of the adjacency matrix indicates the adjacency weights $w_{ij}$ for original graph ${\boldsymbol{A}}_{{\mathcal{G}}}$ and $\widetilde{w}_{ij}$ for sparse graphs ${\boldsymbol{A}}_{{\mathcal{G}_S}}$. The empty white spaces indicate sparsity in the adjacency matrix. Sparsity index is $S \equiv n_{\text{non-zero}}/N^2$, where $n_{\text{non-zero}}$ is the number of non-zero weights (elements) in adjacency matrix and $N$ is the number of vortices. The colored circles in the vortex network represent the individual cluster groups.}
   \label{fig_graph}
\end{figure}

The original point vortex distribution modeled as a complete graph is shown in figure \ref{fig_graph} (top left). The colors of the point vortices (network vertices) indicate their unweighted degree (number of edge connections). As the original graph here is a complete graph $\mathcal{K}_N$, each vortex is connected to every other vortex. Thus, the unweighted degree of each vortex in this example is $N-1 = 99$. The sparsity pattern of the adjacency matrix for the complete graph $\boldsymbol{A}_{{\mathcal{K}_N}}$ is shown in figure \ref{fig_graph} (top right).  We also note in the figure the sparsity index $S$, defined as $S \equiv n_{\text{non-zero}}/N^2$ with $n_{\text{non-zero}}$ being the number of non-zero weights in the adjacency matrix and $N$ being the number of vortices. As the diagonal elements of the adjacency matrix have zero weights, $n_{\text{non-zero}} = N(N-1)$ for the complete graph. For $N = 100$, a sparsity index of $S = 0.99$ is obtained for the original graph.

We now perform spectral sparsification on the original graph ($\mathcal{G} = \mathcal{K}_N$) using the algorithm $\mathcal{G}_S = \tt{Sparsify}(\mathcal{G})$ described in \S{\ref{sec:esparse}} to obtain the spectrally similar sparse representations of the complete graph based on approximation order $\epsilon$. The corresponding adjacency matrix, ${\boldsymbol{A}}_{{\mathcal{G}_S}}$ can also be found. The vortex network for sparse graph and the sparsity pattern of ${\boldsymbol{A}}_{{\mathcal{G}_S}}$ corresponding to approximation order of $\epsilon = 0.5$ are shown in \fig \ref{fig_graph} (middle). We observe that the number of edges (connections) between the vortices in cluster $3$ and $5$ are reduced. Because the weights of the adjacency matrix are inversely proportional to the distance between the vortices, a large number of edges between the vortices of the clusters with larger distances are cut during sparsification. The sparsity index decreases from $S = 0.99$ for the original complete graph to $ S = 0.741$ for the sparse graph ($\epsilon = 0.5$), reducing the number of connections by approximately $25$ percent.

The vortex network and sparsity patterns for approximation order of $\epsilon = 1$ are shown in figure \ref{fig_graph} (bottom). We observe dense representation of the connections between the clusters $1$ and $2$. Thus, the proximity in clusters is identified both in the network structure and the sparsity patterns.  Also, we realize from the sparsity patterns that the connections between the vortices in a particular cluster are maintained while majority of the inter-cluster ties are cut resulting in a sparse graph. The number of connections in the sparse graph for $\epsilon = 1$ are reduced dramatically by nearly $60$ percent with $S = 0.4006$. 

\begin{figure}
   \begin{center}
     \begin{tabular}{cc}
       (a) & (b) \\ 
        \begin{overpic}[width=0.46\textwidth]{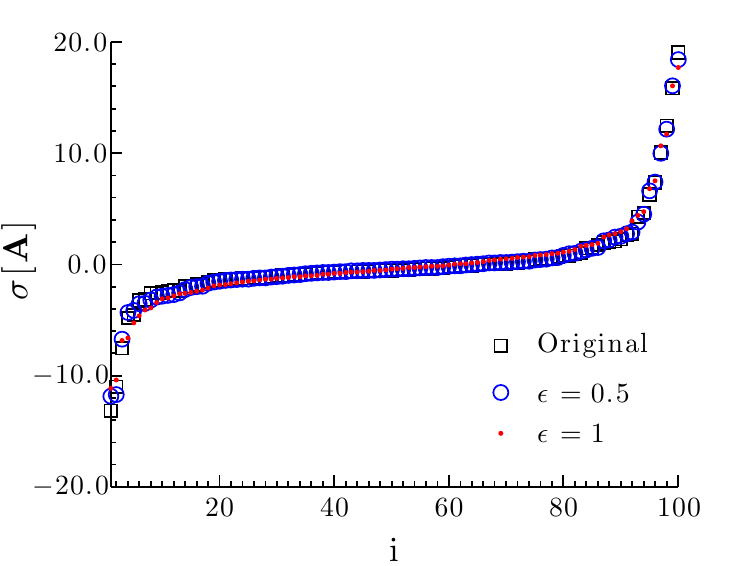}
        \end{overpic} &
        \begin{overpic}[width=0.46\textwidth]{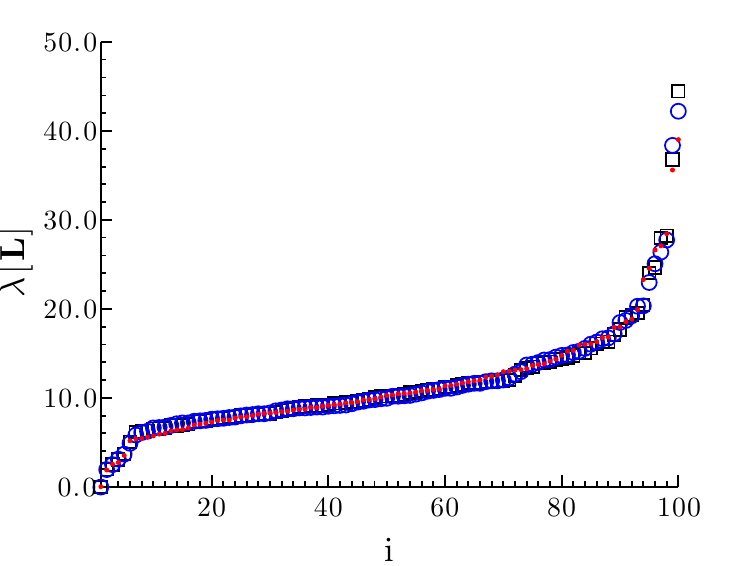}
        \end{overpic} 
      \end{tabular}
   \end{center}
   \caption{Eigenvalue spectra of (a) adjacency and (b) Laplacian matrices of the sparsified graphs with $\epsilon = 0.5$ and $1$ in comparison to the original graph.}
   \label{fig_spectra}
\end{figure}

For weighted graphs, the effective resistance is computed in such as way so as to maintain spectral similarity. The eigenspectra ($\sigma$ and $\lambda$) of the adjacency and Laplacian matrices for the sparse and original graphs are compared in figure \ref{fig_spectra}. We observe from figure \ref{fig_spectra} that the spectra of the sparse and original configurations are in good agreement as expected. The second smallest eigenvalue of the Laplacian matrix represents the spectral gap or algebraic connectivity of the graph, i.e., vertices connected by at least a single path \citep{Newman10}. As observed from \fig{\ref{fig_spectra}}, the spectral gap is greater than zero for original and sparse graphs indicating that the graph is connected. Thus, sparsification preserves the connectivity properties of the original graph. The sparsified graphs slightly underpredict the maximum eigenvalues of the adjacency and Laplacian matrices compared to the original graph. The preservation of spectral properties is attractive if the sparsified network is to be used to describe the dynamics of the vortices. We expect the overall motion of the vortices to remain similar with sparsified representations of the vortex interactions. 

\begin{figure}
   \begin{center}
     \begin{tabular}{cc}
        \begin{overpic}[height=0.4\textwidth]{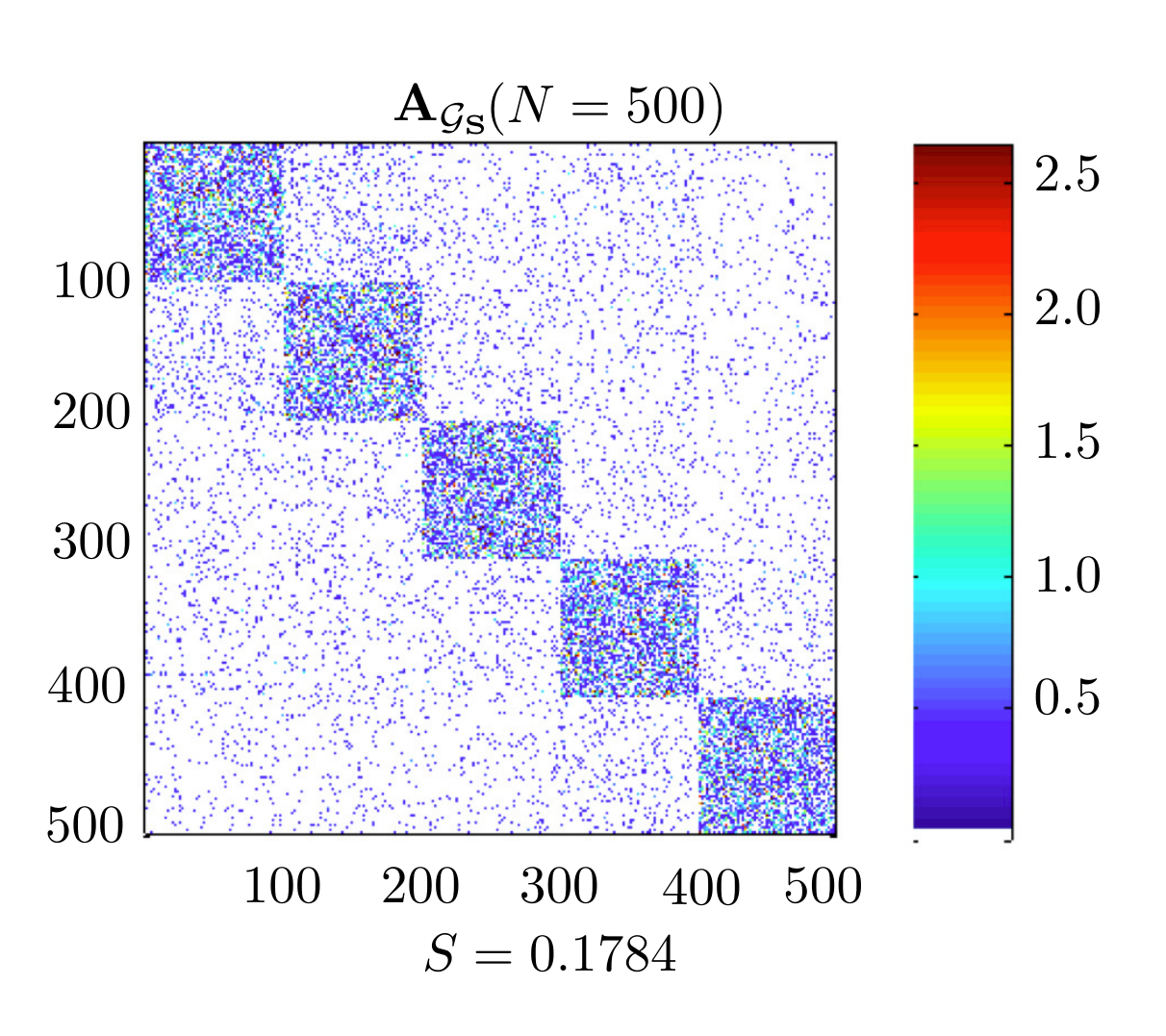}
        \end{overpic} 
        \begin{overpic}[height=0.4\textwidth]{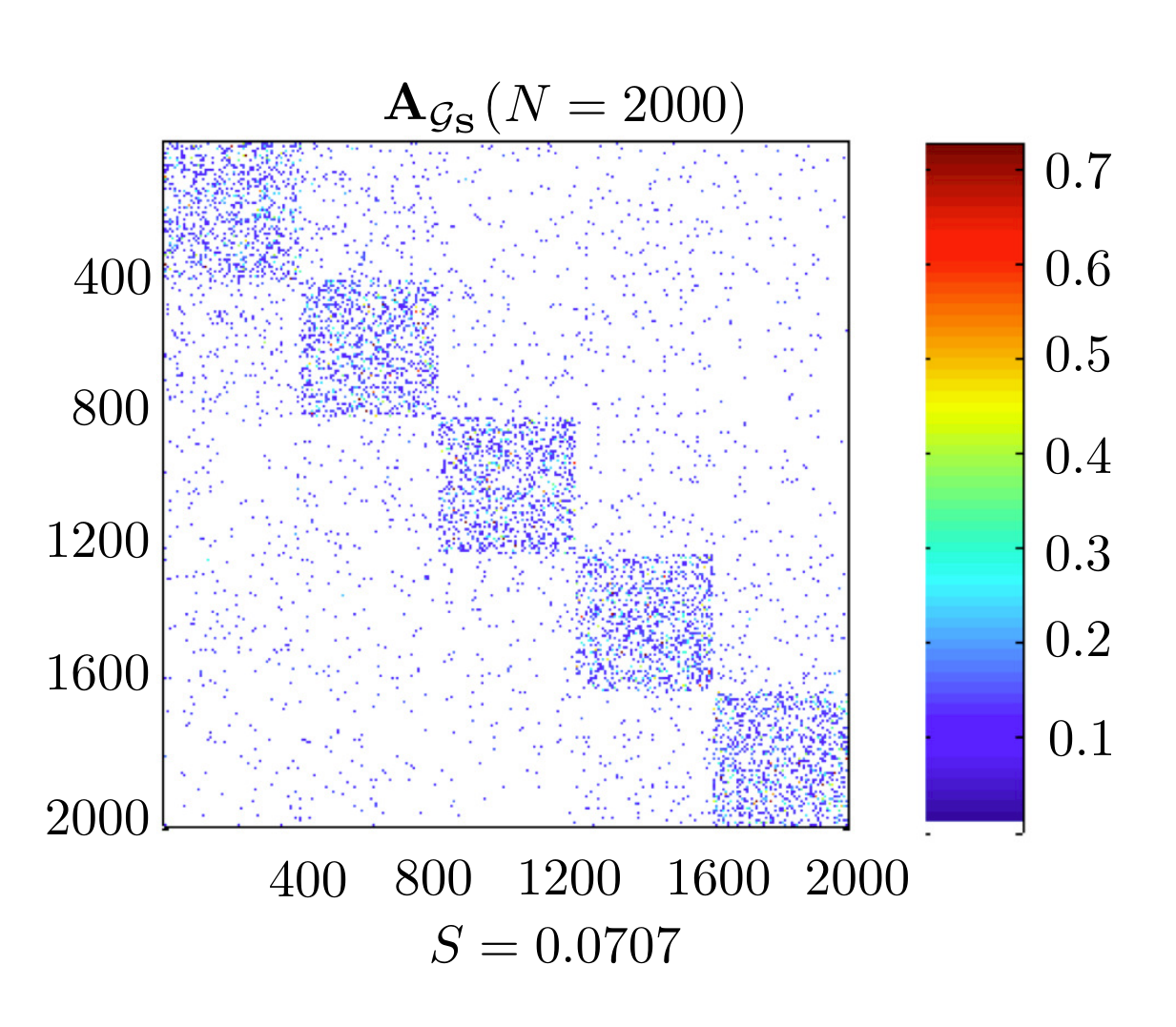}
        \end{overpic} 
   \end{tabular}
   \end{center}
   \caption{Sparsity patterns of adjacency matrix for $N=500$ and $2000$ for approximation order $\epsilon=1$. The color in the sparsity patterns of the adjacency matrix denotes the adjacency weights $\widetilde{w}_{ij}$ with the empty white space indicating sparsity.}
   \label{fig_spapat}
\end{figure}

\begin{figure}
   \begin{center}
     \begin{tabular}{c}
        \begin{overpic}[height=0.4\textwidth]{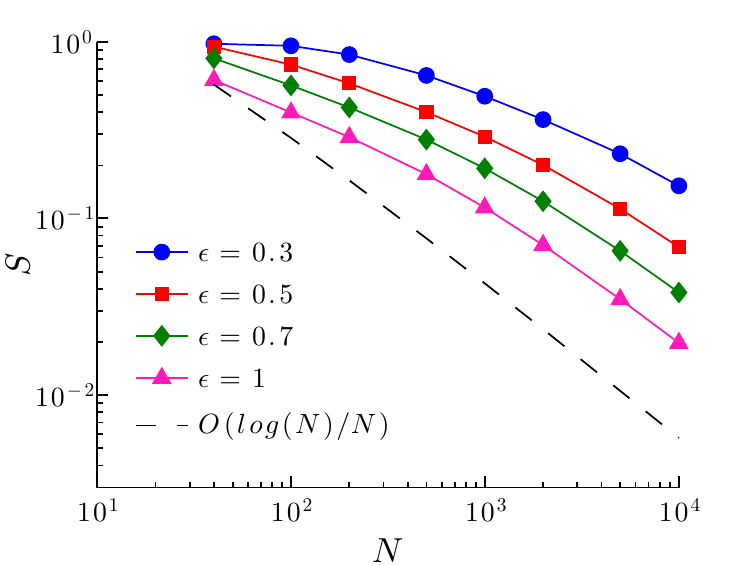}
        \end{overpic} 
   \end{tabular}
   \end{center}
   \caption{Variation of sparsity index $S$ for sparsification with increasing number of point vortices. The reference black line represents $\log(N)/N$.}
   \label{fig_conv}
\end{figure}

Sparsification helps identify the individual clusters by creating subgraphs within a graph where the density of the edges between vertices is much greater than that outside it. The spectral sparsification procedure leads to the reduction in the number of edges to $\mathcal{O}(N\log (N)/\epsilon^2)$ for large $N$. We examine the performance of spectral sparsification on the current example for increasing the total number of vortices. As in the example that was previously considered, we maintain a constant number of clusters $n_c = 5$ but increase the number of vortices in each cluster $N/n_c$. The sparsity patterns of the adjacency matrix for $N = 500$ and $2000$ with approximation order $\epsilon = 1$ are shown in \fig{\ref{fig_spapat}}. We observe that as the number of vortices increases, the adjacency matrix of the sparsified graph becomes increasingly sparse. Similar to the case with $N = 100$ vortices, the majority of the ties within a cluster are maintained while a large number of inter-cluster ties are cut. The sparsity index decreases from $S = 0.1784$ to $0.0707$ for $N = 500$ to $2000$, which is a substantial amount of sparsification. Let us further show the sparsity index $S$ for increasing $N$ with different approximation order $\epsilon$ in \fig \ref{fig_conv}. The sparsity index $S$ decreases with an increase in $\epsilon$ and the number of vortices. The expected behavior of the sparsity index $S = \mathcal{O}(N \log(N))/\mathcal(N^2) = \mathcal{O}(\log (N)/N)$ is observed for larger $N$. The trend deviates from the expected behavior for lower $N$ as the availability of the connections for redistribution of the weights of the sparsified connections decreases. 

We observe that the sparsification algorithm provides us with a heavily sparsified model to produce a computationally tractable representation of the vortex-to-vortex interaction. We further note that the sparsification algorithm can be easily parallelized, which is attractive to further reduce the computational wait time. The bulk motion of clusters of point vortices could be tracked effectively by tracking their individual centroids. The present approach may lead to algorithms similar to fast particle summation methods where near and far-field effects are taken into consideration \citep{Greengard:JCP87}. 

%%%%%%%%%%%%%%%%%%%%%%%%%%%%%%%%%%%%%%%%

\subsection{Sparsified-dynamics model}
\label{sec:svdy}

The sparsified vortex interactions can be used for analyzing the dynamics of the system of discrete point vortices. Spectral sparsification can be viewed as adjusting the strength of the vortices in the Biot-Savart law through the sparsification factor $W$ given by equation (\ref{eq6}). We can incorporate the weights of the sparse graph into the Biot-Savart law as
\begin{align}
 &\frac{d \boldsymbol{r_i}}{d t} = 
 \sum_{\substack{j=1\\ W_{ij}\ne 0}}^N W_{ij} \frac{\kappa_j}{2\pi} \frac{\widehat{\boldsymbol{k}} \times 
 (\boldsymbol{r}_i - \boldsymbol{r}_j)}{|\boldsymbol{r}_i - \boldsymbol{r}_j|^2}, \label{eq14}
\end{align}
where $W_{ij}$ is the sparsification factor. The above expression significantly reduces the amount of computation compared to the original Biot--Savart law, because the sparsification has made a significant number of elements of $W$ to be zero, as seen in figures \ref{fig_graph} and \ref{fig_spapat}.

\begin{figure}
   \begin{center}
     \begin{tabular}{c|cc}
       & \bf{\hspace{0.08\textwidth} \parbox{5cm}{\begin{center} Trace of cluster centroids \\ for $\epsilon = 0.5$ \end{center}}} &
           \bf{\hspace{0.08\textwidth} \parbox{5cm}{\begin{center} Error in dynamics \\ for $\epsilon = 0.5$ \end{center}}}               \\ \hline
        \begin{sideways} \hspace{0.08\textwidth} \parbox{3.8cm}{(a) One time sparsification} \end{sideways} &
        \includegraphics[width=0.48\textwidth]{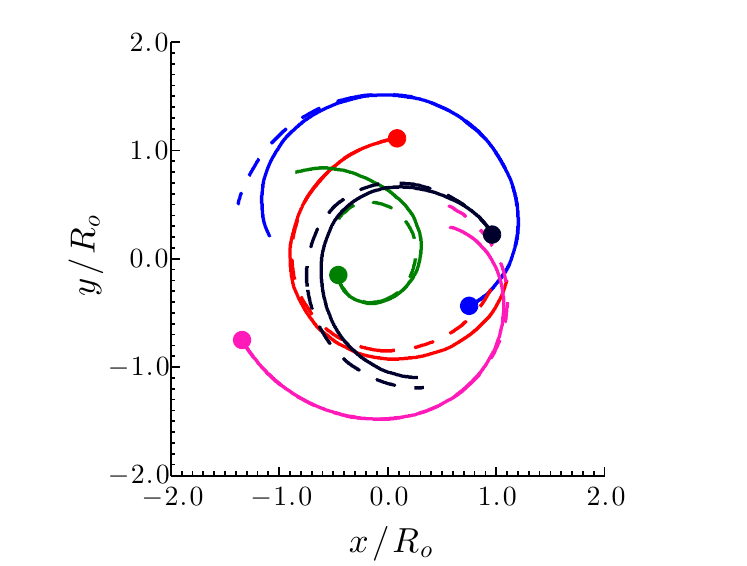} &
        \includegraphics[width=0.48\textwidth]{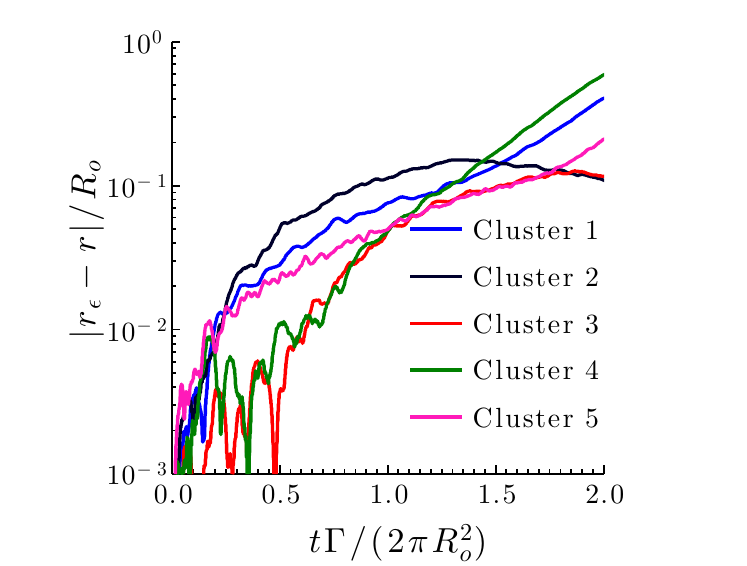} \\ 
        \begin{sideways}\hspace{0.1\textwidth}\parbox{3.5cm}{\begin{center}(b) Resparsification \\ every ($0.1t\Gamma/2\pi R_o^2$) \end{center}}\end{sideways}&
        \includegraphics[width=0.48\textwidth]{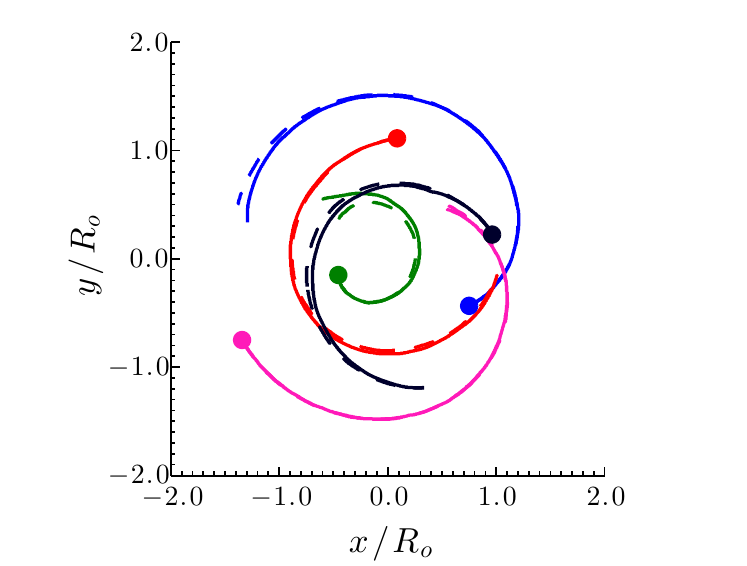} &
        \includegraphics[width=0.48\textwidth]{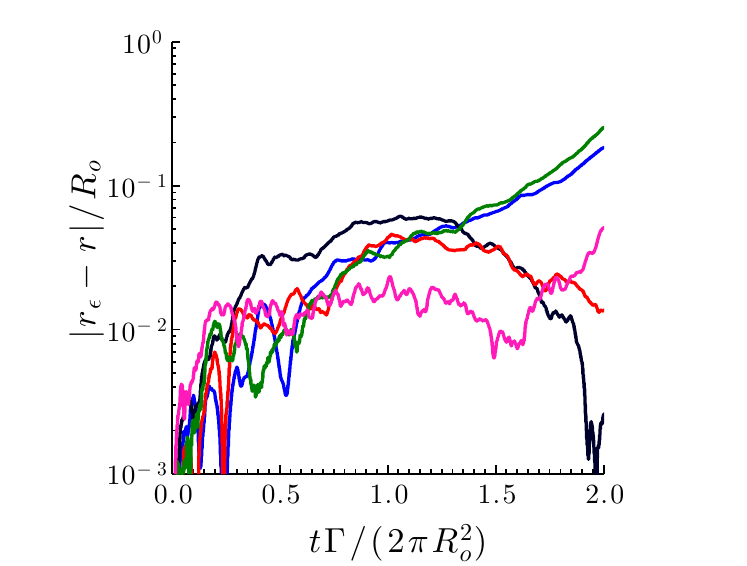} \\
         \begin{sideways}\hspace{0.1\textwidth}\parbox{3.5cm}{\begin{center} (c) Resparsification \\ every ($0.01t\Gamma/2\pi R_o^2$) \end{center}}\end{sideways}&
        \includegraphics[width=0.48\textwidth]{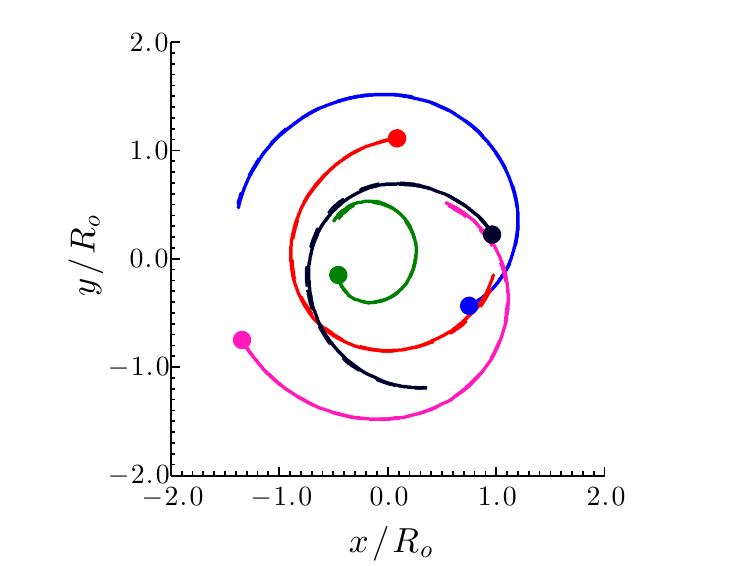} &
        \includegraphics[width=0.48\textwidth]{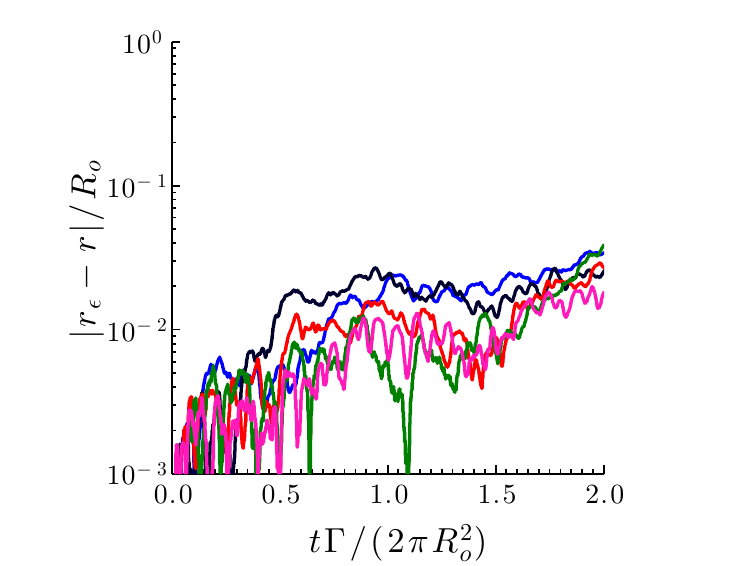} 
        \end{tabular}
   \end{center}
   \caption{Trajectories of the centroids (left) and error in position of centroids (right) of the vortex clusters based on sparse graph ($\epsilon = 0.5$) and original graph with time for $N = 100, \bar{\kappa} = 0.1$. (a) trajectories and error based on sparsification performed only at the initial step while (b) results based on resparsification performed at every ($0.1t\Gamma/2\pi R_o^2$) and (c) results based on resparsification performed at every ($0.01t\Gamma/2\pi R_o^2$).  The full nonlinear solution is shown with dashed lines.}
   \label{fig_d}
   \vspace{-2mm}
\end{figure}

\begin{figure}
   \begin{center}
     \begin{tabular}{c|cc}
       & \bf{\hspace{0.08\textwidth} \parbox{5cm}{\begin{center} Trace of cluster centroids \\ for $\epsilon = 1$ \end{center}}} &
           \bf{\hspace{0.08\textwidth} \parbox{5cm}{\begin{center} Error in dynamics \\ for $\epsilon = 1$ \end{center}}}               \\ \hline
        \begin{sideways} \hspace{0.08\textwidth} \parbox{3.8cm}{(a) One time sparsification} \end{sideways} &
        \includegraphics[width=0.48\textwidth]{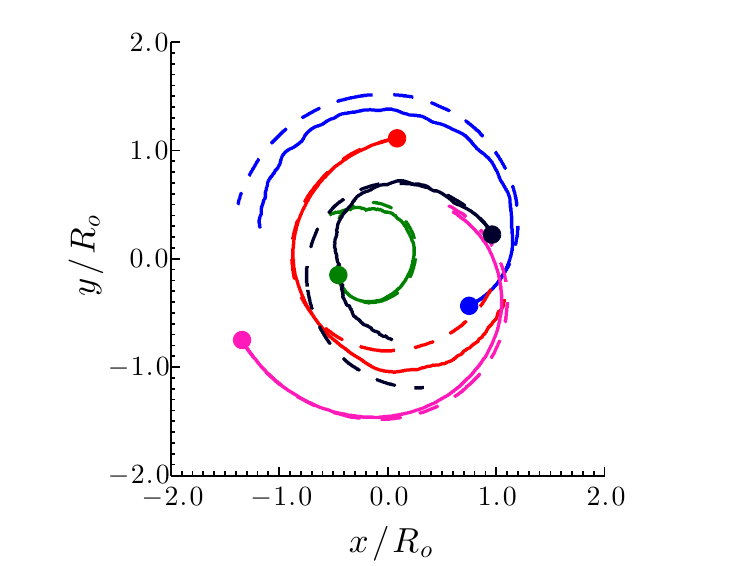} &
        \includegraphics[width=0.48\textwidth]{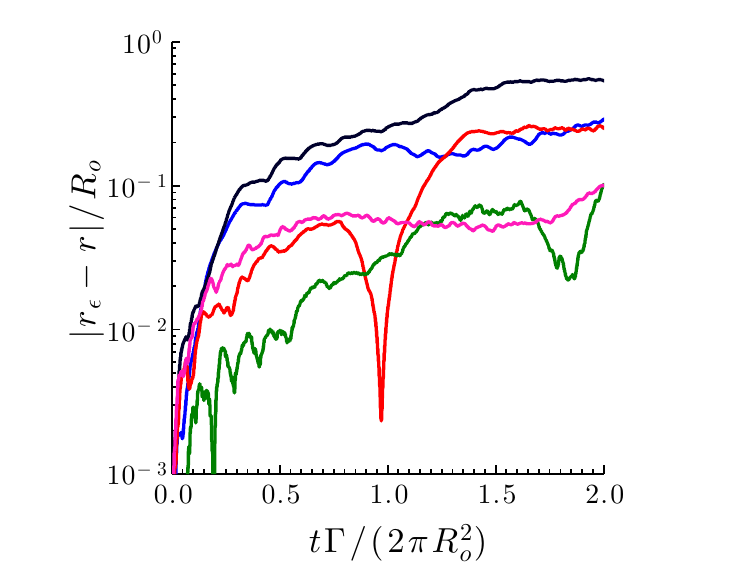} \\ 
        \begin{sideways}\hspace{0.1\textwidth}\parbox{3.5cm}{\begin{center} (b) Resparsification \\ every ($0.1t\Gamma/2\pi R_o^2$) \end{center}}\end{sideways}&
        \includegraphics[width=0.48\textwidth]{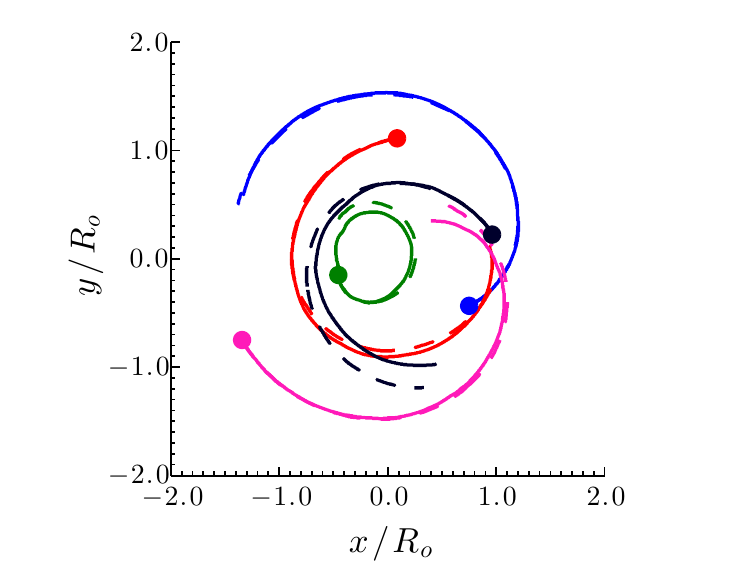} &
        \includegraphics[width=0.48\textwidth]{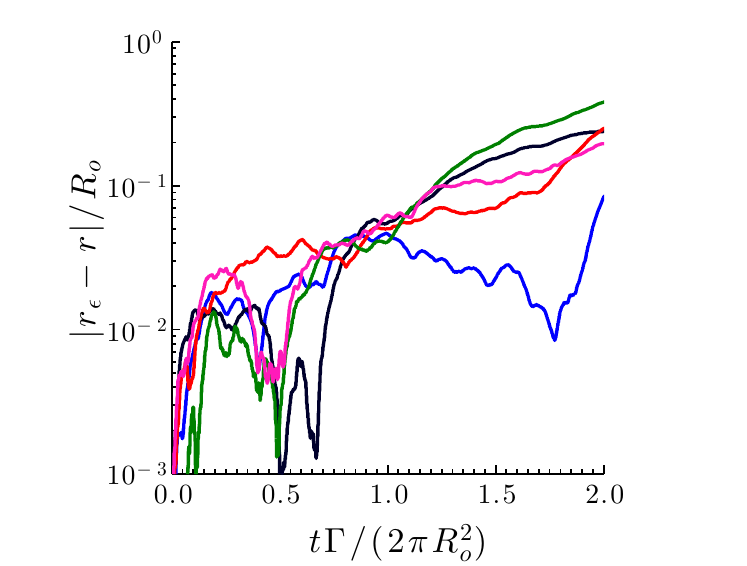} \\
         \begin{sideways}\hspace{0.1\textwidth}\parbox{3.5cm}{\begin{center} (c) Resparsification \\ every ($0.01t\Gamma/2\pi R_o^2$) \end{center}}\end{sideways}&
        \includegraphics[width=0.48\textwidth]{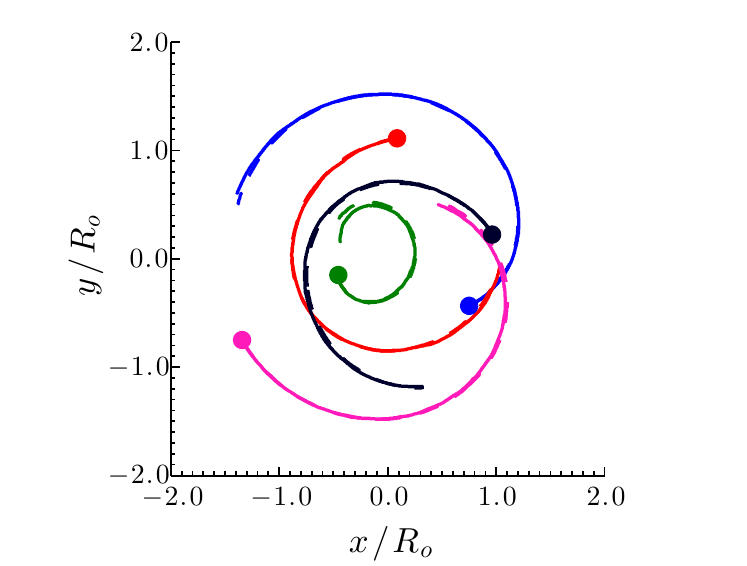} &
        \includegraphics[width=0.48\textwidth]{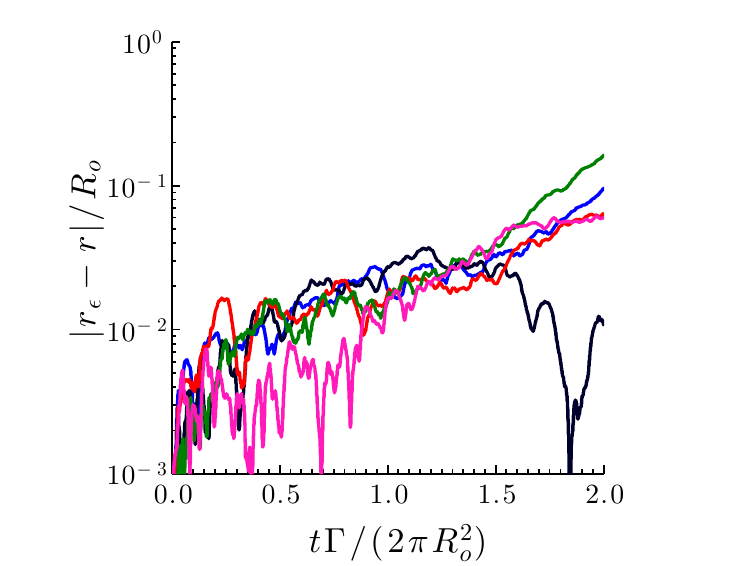} 
        \end{tabular}
   \end{center}
  \caption{Trajectories of the centroids (left) and error in position of centroids (right) of the vortex clusters based on sparse graph ($\epsilon = 1$) and original graph with time for $N = 100, \bar{\kappa} = 0.1$. (a) trajectories and error based on sparsification performed only at the initial step while (b) results based on resparsification performed at every ($0.1t\Gamma/2\pi R_o^2$) and (c) results based on resparsification performed at every ($0.01t\Gamma/2\pi R_o^2$).  The full nonlinear solution is shown with dashed lines.}
   \label{fig_d1}
   \vspace{-2mm}
\end{figure}

For evaluating the effect of sparsification on dynamics of the point vortices, we consider the same example presented in \S{\ref{sec:rom}} with $N=100$ vortices as the initial condition. The objective here is to accurately capture the dynamics of the vortex clusters with the sparsified Biot--Savart law given by equation (\ref{eq14}). The bulk motions of the clusters of point vortices are well represented by the centroid of the individual clusters. The original and the sparsified dynamics are given by equations (\ref{eq53}) and (\ref{eq14}), respectively, and are integrated in time with the fourth-order Runge--Kutta method.  The results from the sparsified-dynamics model are shown in figures \ref{fig_d} and \ref{fig_d1}.

For the present analysis, time is non-dimensionalized by considering the cumulative circulation of the vortices of the cluster ($\Gamma$) and average radial distance of the centroid of the clusters from the geometric center of the overall system at the initial time ($R_o$). As the advective velocity of an isolated cluster of point vortices of strength $\Gamma$ at $R_o$ is given by $u^\star = \Gamma/2\pi R_o$, the non-dimensional time can be deduced as $tu^\star/R_o = t\Gamma/2\pi R_o^2$. In the current example, each individual cluster has a collection of vortices with their strength having a mean of $\bar{\kappa} = 0.1$ (with a normal distribution). Thus, for $n$ vortices in a cluster, $\Gamma = n\bar{\kappa}$. The error in position of the centroid of the clusters for the sparsified setting with respect to the original setting given by $|\boldsymbol{r}_\epsilon - \boldsymbol{r}|$ is non-dimensionalized by the average radial distance of the centroid of the clusters from the geometric center of the system at the initial time ($R_o$). 

Let us consider the case where sparsification is performed only once to determine the sparsification factor $W_{ij}$ before initiating the time integration. The same sparsification factor is used throughout the time integration. This procedure is inexpensive as sparsification is performed only once {\it a priori}. The trace of the positions of the centroid of the individual clusters and their error for approximation orders of $\epsilon=0.5$ and $1$ for a single sparsification are shown in figures \ref{fig_d} (top) and \ref{fig_d1} (top), respectively. Considering the number of connections cut between the vortices for sparse approximations, the trajectories of the centroid of the clusters based on sparse vortex dynamics agrees reasonably with those from full dynamics. On comparing the errors, we find that the errors with approximation order of $\epsilon=0.5$ is less than those with $\epsilon=1$, which is expected.  Despite the Biot--Savart law being nonlinear, the present sparsified approach achieves a reasonable agreement with the full nonlinear solution.

If the clusters of vortices are far from each other, the connections cut between their vortices do not cause a large change in the dynamics while the connections cut between clusters close to each other affect the dynamics considerably. As time progresses, the relative distance between vortices change, resulting in an increase in error. For $t\Gamma/2\pi R_o^2 > 1$, we observe that the errors in sparse vortex dynamics increases to $\approx 0.5 R_o$.  As the adjacency weights $w_{ij}$ and the sparsification factor $W_{ij}$ are based on the initial position of the vortices, the ratio does not hold over large times. This calls for adjustment of the weights and periodic sparsification to adapt to the dynamically changing position of the vortices. 

We thus consider resparsification of the graph representing the system of point vortices periodically at ($0.1t\Gamma/2\pi R_o^2$) and ($0.01t\Gamma/2\pi R_o^2$) for both approximation orders. The trace of the centroids and error for approximation orders of $\epsilon = 0.5$ and $1$ with resparsification are shown in middle and bottom subfigures, respectively, of figures \ref{fig_d} and \ref{fig_d1}. We observe that the error decreases significantly with resparsification. The error in position reduces from $\mathcal{O}(10^{-1})$ to $\mathcal{O}(10^{-2})$ with resparsification. We notice that the centroid trajectories based on the sparse and original calculations become increasingly similar with resparsification. Resparsification updates the sparsification factor periodically based on the position and strength of the vortices and decreases the error by an order of magnitude. Thus, the nonlinear evolution of discrete vortex dynamics is well predicted by the sparsification techniques. 

\begin{figure}
   \begin{center}
  \begin{tabular}{cc}
       (a) & (b) \\
        \includegraphics[width=0.45\textwidth]{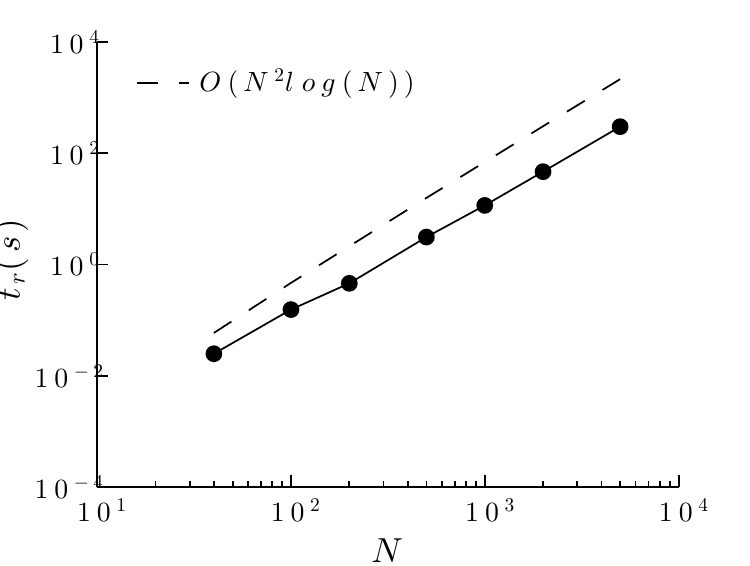} &
        \includegraphics[width=0.45\textwidth]{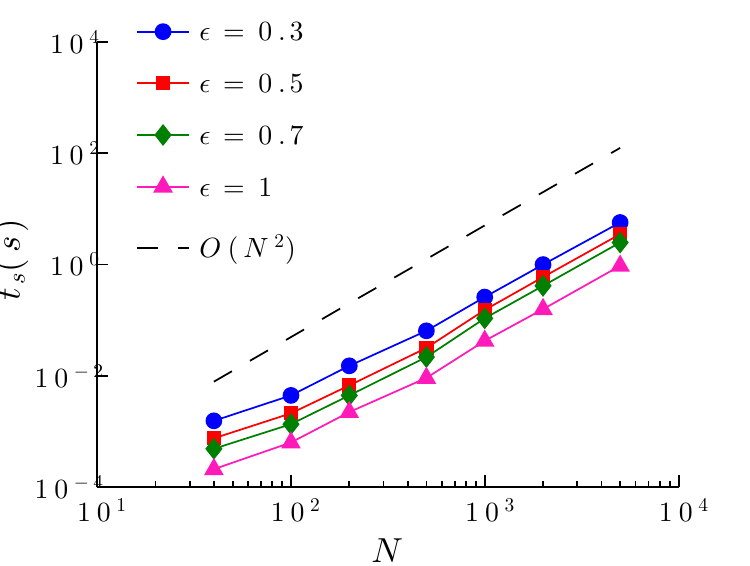} \\
        \multicolumn{2}{c}{(c)} \\ 
        \multicolumn{2}{c}{\includegraphics[width=0.45\textwidth]{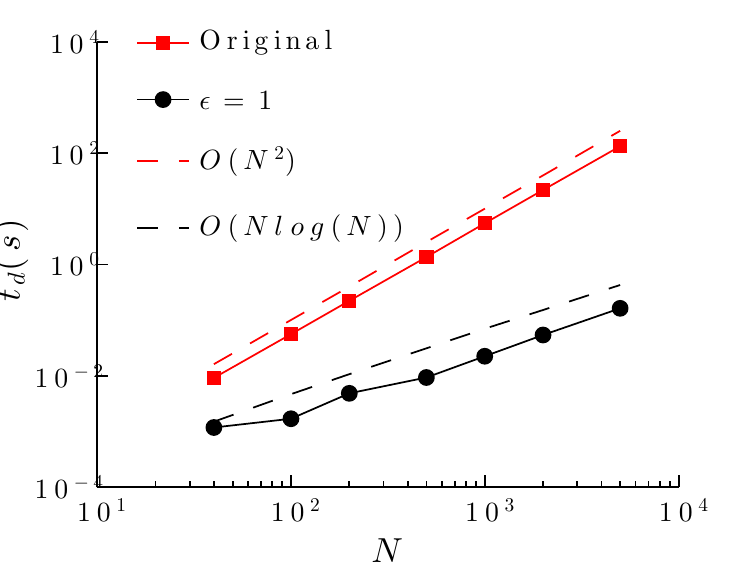}}
      \end{tabular}
   \end{center}
   \caption{The time required for computation of (a) effective resistance and (b) random sampling for different approximation orders for one-time sparsification. The dashed lines indicate the expected trends. (c) Time required for numerical integration at each step for original and sparsified configuration with $\epsilon = 1$.}
   \label{fig_timing_conv}
\end{figure}

One of the advantages of sparsification is the decreased computational cost due to increased sparsity of the connections between the vortices. This could potentially lead to design of faster algorithms based on sparsification strategies. Let us first evaluate the offline cost of computing effective resistance and random sampling required for spectral sparsification. The time required for computing effective resistance $t_r$ and time required for random sampling $t_s$ is shown in figures \ref{fig_timing_conv}(a) and (b) respectively. All computations were performed in MATLAB on an iMac with $3.4$ GHz Intel Core i7 processor. We can observe that for each edge, the computation time of effective resistance is $\mathcal{O}(\log(N))$, hence requiring $\mathcal{O}(N^2 \log (N))$ for the overall effective resistance computation. The random sampling procedure takes $\mathcal{O}(N^2)$ time for all the approximation orders. We compare the computational time $t_d$ required at each step of numerical integration of the Biot--Savart law for the original and sparse configuration. We can observe from figure \ref{fig_timing_conv}(c) that the original configuration takes $\mathcal{O}(N^2)$ time equivalent to the number of edges in the complete graph, while for the sparse configuration the computation time is reduced to $\mathcal{O}(N \log (N))$.

\begin{figure}
   \begin{center}
     \begin{tabular}{c}
        \includegraphics[width=0.5\textwidth]{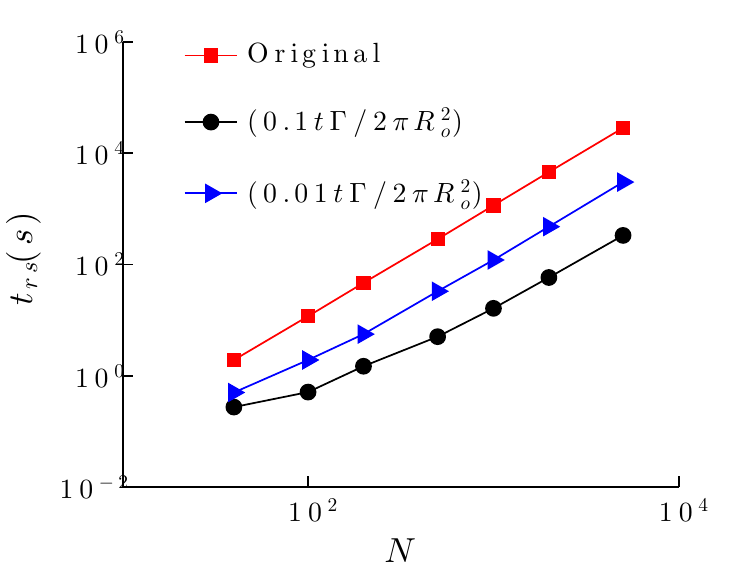} 
      \end{tabular}
   \end{center}
   \caption{The time required for numerical integration for a total time of $0.1t\Gamma/2\pi R_o^2$ with different resparsification frequencies.}
   \label{fig_resparse_time}
\end{figure}

There is always a tradeoff amongst the level of sparsification, the error that appears in the dynamics, and the associated computational cost. The sparser the network is, the faster the computation can become but may compromise accuracy. To increase the computational accuracy, resparsification can be performed but can introduce an additional computational load. We evaluate the computational cost for numerical integration until a total time of $0.1t\Gamma/2\pi R_o^2$. The time required for resparsification and numerical integration for the two different resparsification frequencies (performed every $0.1t\Gamma/2\pi R_o^2$ and $0.01t\Gamma/2\pi R_o^2$) considered in this work for original and sparse configuration with $\epsilon = 1$ is shown in figure \ref{fig_resparse_time}. It should be noted that for the resparsification performed every $0.1t\Gamma/2\pi R_o^2$, resparsification is performed once for the total time considered. For resparsification conducted every $0.01t\Gamma/2\pi R_o^2$, resparsfication is performed ten times. We can see that the time required for resparsification at every $0.1t\Gamma/2\pi R_o^2$ is considerably less compared to the original configuration. With increase in resparsification frequency, though the time required is less compared to the original configuration, there is increased cost associated with resparsification. 

\begin{figure}
   \begin{center}
     \begin{tabular}{cc}
       (a) & (b) \\
        \includegraphics[width=0.45\textwidth]{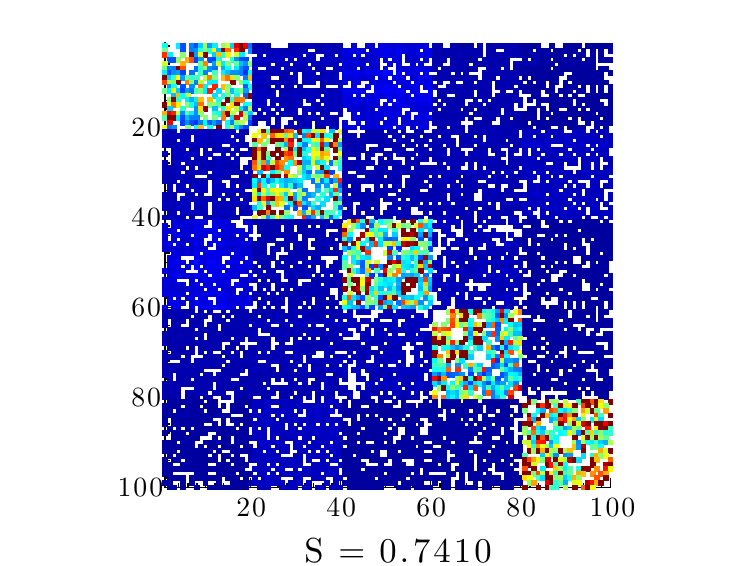} &
        \includegraphics[width=0.45\textwidth]{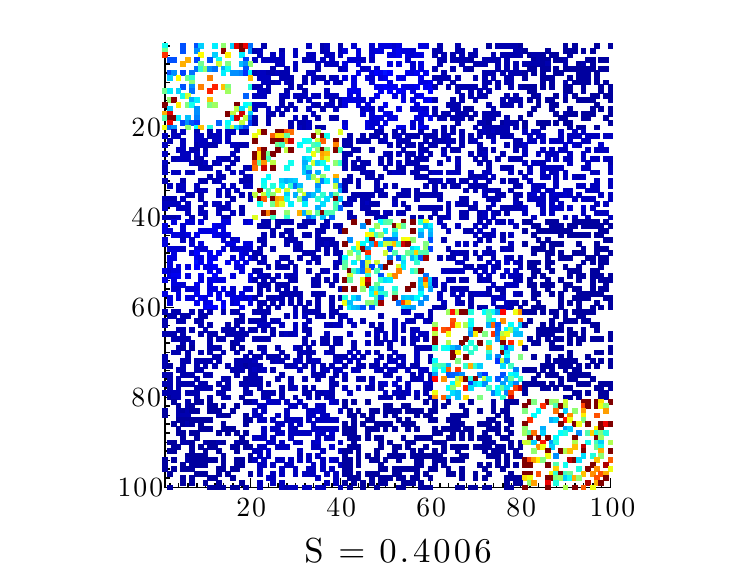} \\
        \multicolumn{2}{c}{(c)} \\ 
        \multicolumn{2}{c}{\includegraphics[width=0.46\textwidth]{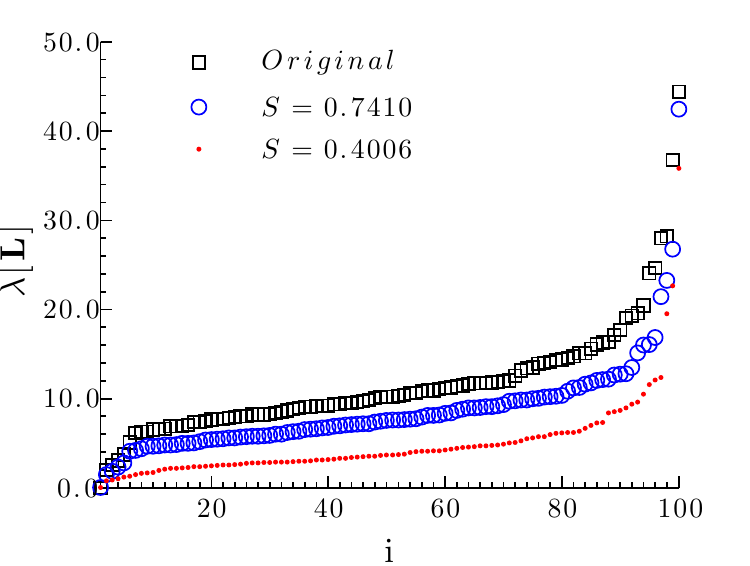}}
      \end{tabular}
   \end{center}
   \caption{Sparsity patterns of the adjacency matrix for $N = 100$ vortices with random cuts with sparsity (a) $S = 0.7410$ and (b) $S = 0.4006$. The color in the sparsity patterns of the adjacency matrix indicates the adjacency weights $\widetilde{w}_{ij}$ for $S = 0.7410$ and $0.4006$. (c) Eigenvalue spectra of Laplacian matrices of the randomly sparsified graphs with $S = 0.7410$ and $0.4006$ in comparison to the original graph. The legend for the sparsity patterns is similar to that of sparsity patterns in figure \ref{fig_graph}.}
   \label{fig_rev1}
   \vspace{-5 mm}
\end{figure}

To illustrate that the trajectory prediction using sparsification cannot be performed naively, we compare the results from spectral sparsification with those based on random edge removal (without redistribution of edge weights).  We keeep the number of cuts identical to that achieved by spectral sparsification for fairness. The sparsity patterns for random cuts with sparsity $S = 0.7410$ and $0.4006$ and the corresponding Laplacian eigenspectra are shown in figure \ref{fig_rev1}. We observe that the spectra from the randomly sparsified graphs grossly under-predicts that of the original graph.

\begin{figure}
   \begin{center}
     \begin{tabular}{cc}
       (a) & (b) \\ 
        \begin{overpic}[width=0.46\textwidth]{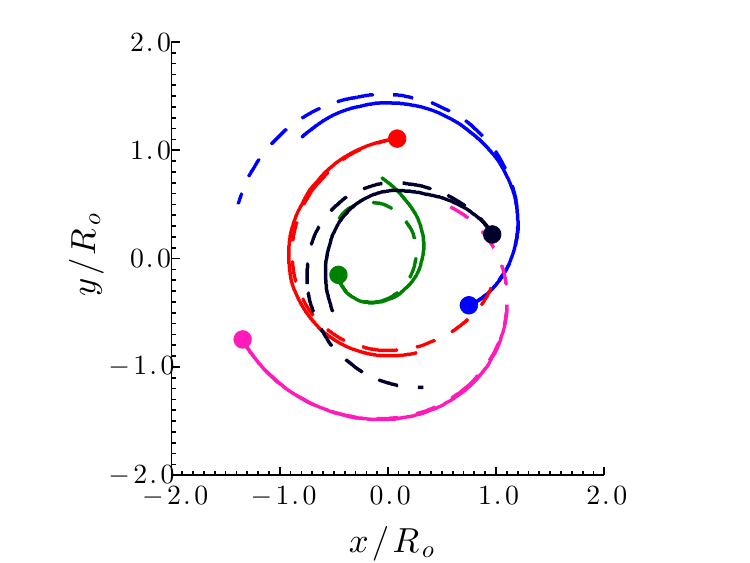}
        \put(-14,38){\rotatebox[]{90}{$S= 0.7410$}}
        \end{overpic} &
        \begin{overpic}[width=0.46\textwidth]{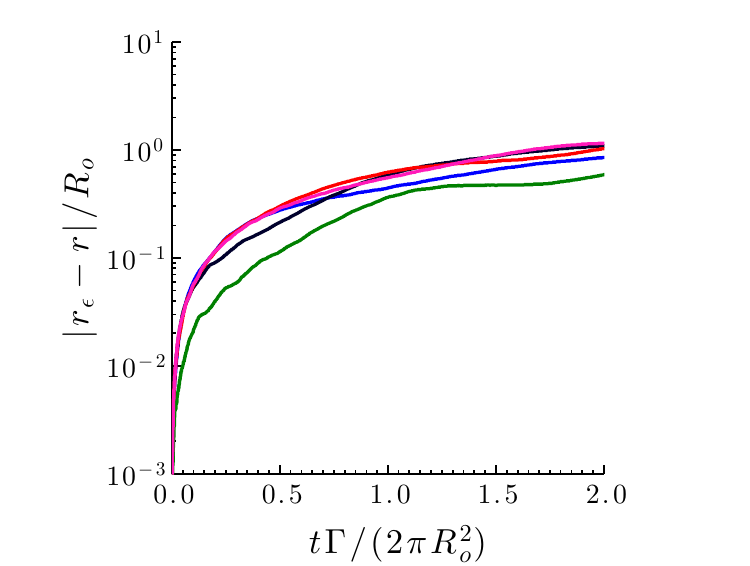}
        \end{overpic} \\
         \begin{overpic}[width=0.46\textwidth]{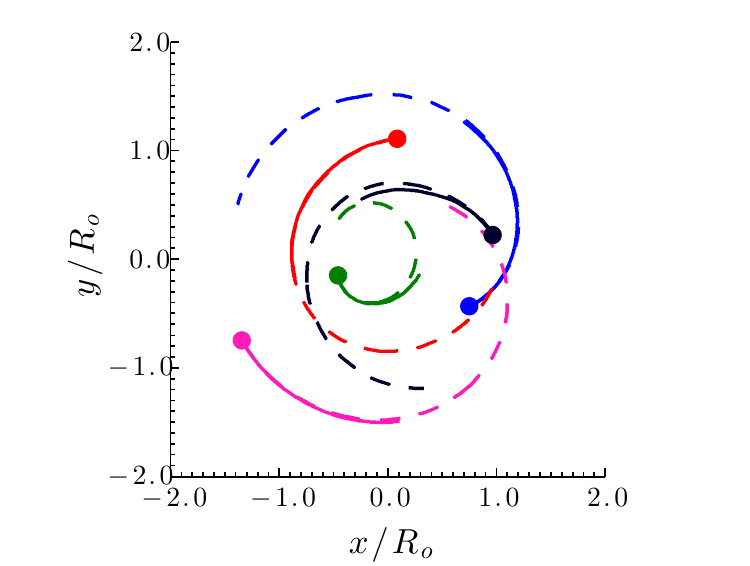}
         \put(-14,38){\rotatebox[]{90}{$S= 0.4006$}}
        \end{overpic} &
        \begin{overpic}[width=0.46\textwidth]{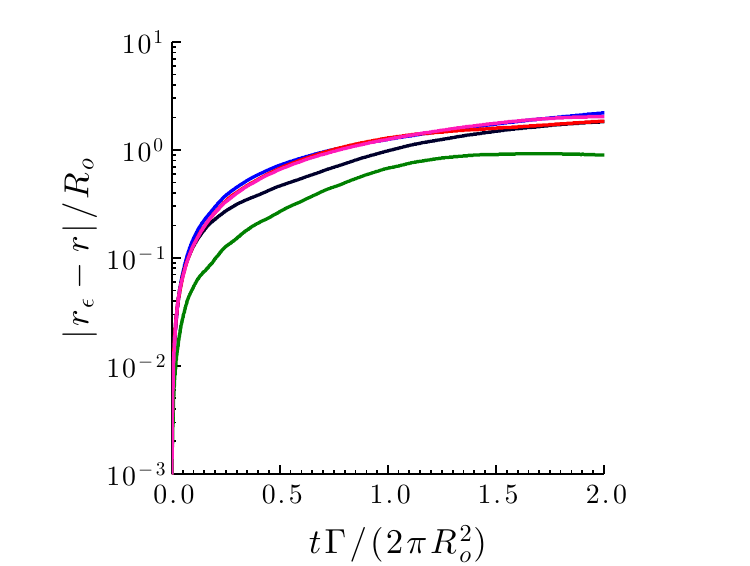}
        \end{overpic} 
      \end{tabular}
   \end{center}
   \caption{(a) Trajectories of the centroids and (b) error in position of centroids (right) of the vortex clusters based on random cuts with sparsity $S= 0.7410$ and $0.4006$ for $N = 100, \bar{\kappa} = 0.1$.}
   \label{fig_rev3}
\end{figure}

The trace of the positions of the centroid of the individual clusters and their error for random cuts with sparsity $S = 0.7410$ and $ 0.4006$ for one-time sparsification are shown in figure \ref{fig_rev3}. The random-cut approximation performs poorly as compared to spectral sparsification and the dynamics of the centroid of the vortex clusters are not captured. As the connections between the vortices are randomly cut, the centroids of the clusters move slower compared to the original configuration as expected. This is due to the loss of induced velocity with the absence of weight redistribution from random sparsification.  Spectral sparsification, on the other hand, redistributes the weights to prevent the loss of the overall interactions amongst the set of point vortices.

It is well known that fast multipole methods \citep{Greengard:JCP87} can also reduce the $\mathcal{O}(N^2)$ velocity evaluation significantly. The particle-box and box-box schemes reduce the computational complexity from $\mathcal{O}(N^2)$ to $\mathcal{O}(N \log(N))$ and $\mathcal{O}(N)$, respectively.  As seen in this section, spectral sparsification reduces the computational load to $\mathcal{O}(N\log(N))$ for computing the velocity of the vortices. Fast multipole methods approximate the effect of cluster of particles at a certain distance by multipole expansions and organizing the particles to a hierarchy of clusters \citep{Cottet00}. In order to compensate for the sparsified connections in spectral sparsification, the weights are redistributed among the other edge connections to compute the sparsified dynamics. In an analogy to fast multipole methods, the interaction list depends on the connections that have direct impact on the vortices and computational accuracy may degrade when the vortices move in space. Similar to reconstruction of tree data structures in fast multipole methods, performing resparsification periodically increases the accuracy by re-evaluating the associated weights as the positions of the vortices evolve over time. We note that even with sparsification, invariants of discrete vortices are conserved as discussed below.

%%%%%%%%%%%%%%%%%%%%%%%%%%%%%%%%%%%%%
\subsection{Conservation properties}

For a discrete set of point vortices, angular impulse and linear impulse are among the conserved quantities. The coordinates of center of vorticity obtained from linear impulse and the length of dispersion about the center of vorticity obtained from angular impulse, when the strength of vortices is of the same sign, remain constant. The center of vorticity $(X,Y)$ and length of dispersion of vorticity $D$ are given by 
\begin{align}
X &= \frac{\sum_{i=1}^N \kappa_i x_i}{\sum_{j=1}^N \kappa_j}, \quad
Y  = \frac{\sum_{i=1}^N \kappa_i y_i}{\sum_{j=1}^N \kappa_j}, \\
D^2 &= \frac{\sum_{i=1}^N \kappa_i \{(x_i - X)^2 + (y_i - Y)^2\}}{\sum_{j=1}^N \kappa_j},
\end{align}   
where $\kappa_i$ is the strength of the $i$-th vortex and $(x,y)$ are the vortex positions \citep{batchelor2000introduction,Newton01}. The circulation for $n$ vortices in a cluster is $\Gamma = n\bar{\kappa}$. The total circulation of the system of $n_c = 5$ clusters is given by $\Gamma_t = \sum_{i=1}^N \kappa_i = n_c\Gamma$. Another conserved quantity is the Hamiltonian. The Hamiltonian is defined as the interaction energy of the system of point vortices and is given by
\begin{equation}
H = -\frac{1}{4\pi}\sum_{i,j = 1, i \ne j}^{N} \kappa_i \kappa_j \log\sqrt{(x_i - x_j)^2 + (y_i - y_j)^2}.
\end{equation}
We non-dimensionalize the Hamiltonian using the total circulation of $n_c$ clusters and the average radial distance of centroid of the clusters at initial time $R_o$. 

\begin{figure}
   \begin{center}
     \begin{tabular}{cc}
       (a) & (b) \\ 
        \begin{overpic}[width=0.42\textwidth]{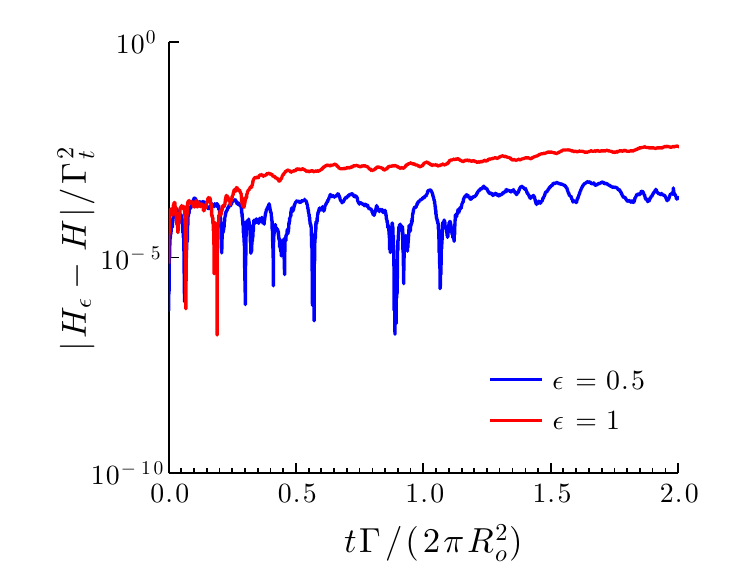}
        \end{overpic} &
        \begin{overpic}[width=0.42\textwidth]{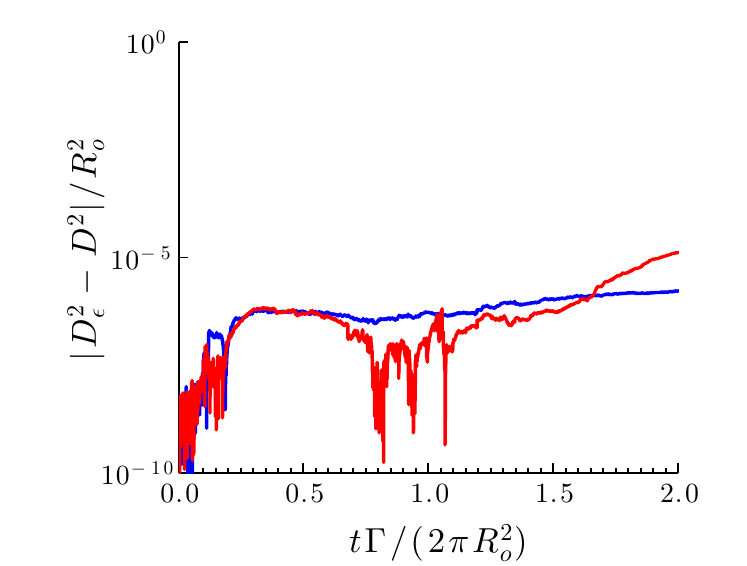}
        \end{overpic} \\
        (c) & (d) \\
        \begin{overpic}[width=0.42\textwidth]{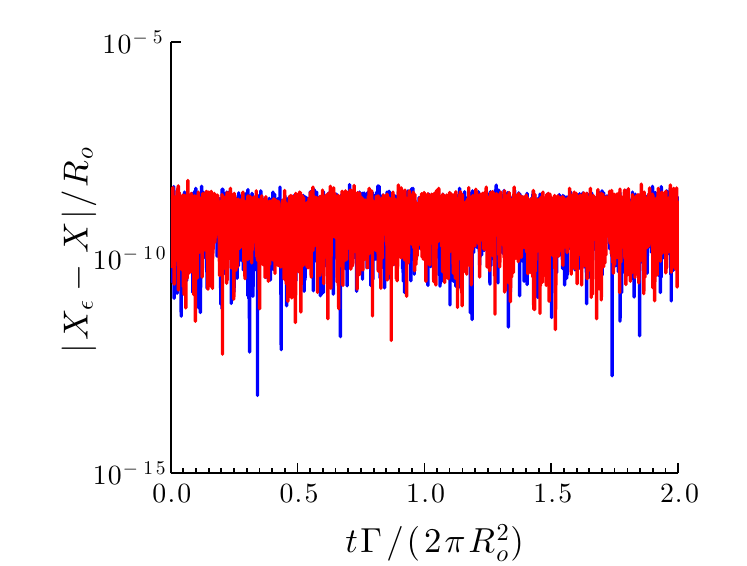}
        \end{overpic} &
        \begin{overpic}[width=0.42\textwidth]{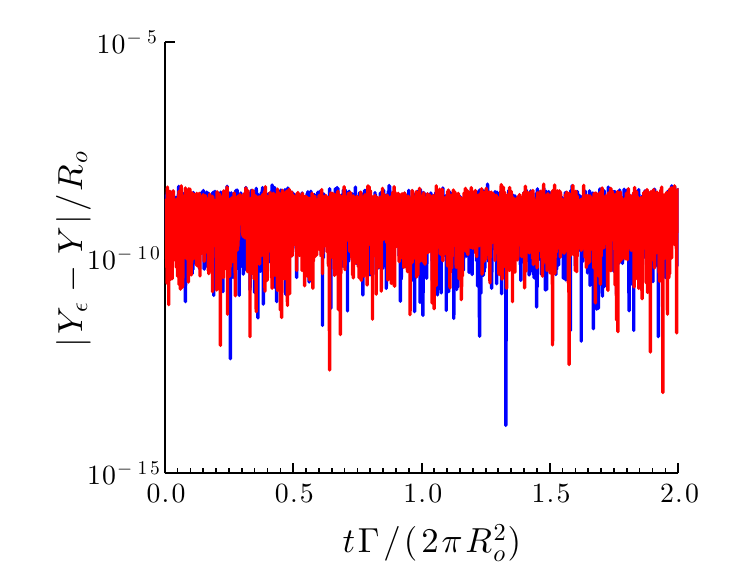}
        \end{overpic} 
     \end{tabular}
   \end{center}
   \caption{Error in the (a) Hamiltonian, (b) square of length of dispersion, (c) $x$-position of center of vorticity and (d) $y$-position of center of vorticity based on single sparsification for $N = 100, \bar{\kappa} = 0.1$ and $\epsilon = \{0.5,1\}$.}
   \label{fig_d2}
\end{figure}

We compare the error in the Hamiltonian $({H}_\epsilon(r/R_o)- H(r/R_o))/\Gamma_t^2$, the center of vorticity $((X_\epsilon - X)/R_o, (Y_\epsilon - Y)/R_o)$, and the square of length of dispersion of vorticity $(D_\epsilon^2 - D^2)/R_o^2$ for approximation orders of $\epsilon = 0.5$ and $1$ over time in figure \ref{fig_d2}. The preservation of center of vorticity and length of dispersion implies the conservation of linear and angular impulse. Here, the variables with subscript $\epsilon$ denote those based on a sparsified model. We observe that the error in the Hamiltonian is of $\mathcal{O}(10^{-4})$ for $\epsilon = 0.5$ and $\mathcal{O}(10^{-3})$ for $\epsilon = 1$. The error in the square of dispersion is of $\mathcal{O}(10^{-6})$ while the error in the center of vorticity is of $\mathcal{O}(10^{-8})$ with sparsification. While not shown, resparsification of these vortices periodically performed at every ($0.1t\Gamma/2\pi R_o^2$) and ($0.01t\Gamma/2\pi R_o^2$) maintain similar error levels for the invariants. We also note that the circulation of each vortices has not been altered, which conserves the individual and overall circulation over time.

As seen in \S\ref{sec:esparse}, sparsification by effective resistance is based on the energy-minimization principle. The effective resistance is defined such that the total energy of the system remains constant. This is reflected well in the conservation of the Hamiltonian representing the interaction energy of the discrete point vortices. In addition, sparsification preserves other invariants including linear and angular impulse of the discrete set of point vortices. We note in passing that even with resparsification, the invariants were observed to be conserved. The conservation of the invariants for the sparse configurations further encourages the applicability of sparsification strategies on discrete vortex dynamics. 

%%%%%%%%%%%%%%%%%%%%%%%%%%%%%%%%%%%%%%%%%%%%%%%%%%%%%%%%
\subsection{Sparsified-dynamics model and Reduced-order model}

We comment on the similarities and differences between the present sparsified-dynamics model and the reduced-order model.  Both models share the general objective of deriving a model that captures the full-order physics in a distilled manner.  Reduced-order models achieve such goals by reducing the dimension of the state variable.  One such technique is the Galerkin projection of the Navier--Stokes equations using a set of spatial bases \citep{Noack:JFM05,Rowley2004model}, such as those determined from POD \citep{berkooz1993proper,Holmes96}.  One can further consider incorporating the effect of input and output dynamics by using balanced truncation or the eigensystem realization algorithm \citep{Rowley:IJBC05, Ma:TCFD09}. The resulting reduced-order models are generally described by ordinary-differential equations for the temporal coefficients with reduced dimensions.

On the other hand, the present sparsified-dynamics model does not reduce the dimensionality of the state variable.  We instead focus on reducing the number of connections used to capture the overall dynamics of the fluid flow.  The reduction in the number of connections is performed based on the concepts from network analysis and graph theory.  Although we cut a large number of the edges in the graph representation of the dynamical interaction, we redistribute the weights associated with the edges to maintain properties of the graph.  This procedure is based upon approximating the spectral properties of the graph and does not require selecting the spatial basis functions unlike the Galerkin-projection based models.  One nice feature of the sparsified-dynamics model is its ability to conserve physical variables such as the Hamiltonian, angular impulse, and linear impulse, as discussed previously. Hence, the sparsified dynamics is able to predict the full dynamics as demonstrated by the example with discrete vortices.  By highlighting the interactions amongst a set of vortices, we are able to determine which interactions amongst the vortices are important in guiding the overall motion of vortices.  

We believe the sparsified-dynamics model has promising potential to model various types of fluid flow by considering the modal structures as a abstraction of graph nodes.  The application of sparsified dynamics models towards flow control problems may also become fruitful as the the computational time necessary to capture the complex behavior of the flow is reduced and the interaction of nodes or flow structures are well-captured, which has been a lacking feature in linear dynamics model and linear stability analysis.  We anticipate that the presently proposed model can lead to potential feedback control of fluid flow  \citep{Bagheri:AMR09, Ahuja:JFM10} but with nonlinear interactions emphasized. One of the open questions in extending the sparsified dynamics model for a wide variety of fluid flow problems is the choice of the variables or modes to be used for graph nodes. This issue is currently being examined and will be reported in upcoming studies.

%%%%%%%%%%%%%%%%%%%%%%%%%%%%%%%%%%%%%%%%%%%%%%%%%%%%%%%%

\section{Concluding remarks}
\label{sec:conc}

Graph representations of a system of discrete point vortices provide us with a framework for identifying important connections amongst vortices. The edge weights of the graph were related to the importance of the ties between the vortices, which were quantified based on the strength of the vortices and the relative distance between them. We sparsify the connections between these vortices based on spectral graph theory not only to cut some of these ties but also to redistribute their weights to preserve the dynamics of the vortices.  We have demonstrated that an effective resistance approach to graph sparsification proves extremely useful in constructing sparsified-dynamics models. Moreover, sparsification reduces the number of connections to $\mathcal{O}(N\log N/\epsilon^2)$ based on approximation order $\epsilon$. We also observe an increase in the effectiveness of sparsification with increasing number of point vortices. In spite of the nonlinear nature of the dynamics of these point vortices, the sparse representations are capable of identifying the core vortex structures as well as tracking the bulk motion of these structures in an effective manner. We have referred to the dynamical model that uses the sparsified vortex network as the sparsified-dynamics model. In addition, the sparsified models preserve spectral properties of the original setup. Resparsification is utilized to enhance the prediction of  the motion of the vortices over time. We observe that sparsification conserves the invariants of discrete vortices dynamics such as the Hamiltonian, linear and angular impulse, and circulation. The objectives of sparsified-dynamics model are aligned with those of reduced-order models but these goals are achieved by sparsifying the connections rather than deriving a model equation with state variable of reduced dimension. The results of graph-theoretic approach to discrete vortex dynamics point to promising research in its applicability to a variety of problems in fluid mechanics.

\section{Acknowledgements}

We thank Dr.~Matthew Jemison and Prof.~Louis Cattafesta for their comments on an earlier version of this paper.  This work was supported by the US Air Force Office of Scientific Research (Grant FA9550-13-1-0183, Program Manager: Dr.~Douglas Smith) and the US Army Research Office (Grant W911NF-14-1-0386, Program Manager: Dr.~Samuel Stanton).  AGN also acknowledges the Aeropropulsion, Mechatronics, and Energy fellowship from the Florida State University.
%=====================================================================

%=====================================================================

\bibliographystyle{jfm}
\bibliography{vortex_network} % change as needed (file directory)

\begin{thebibliography}{41}
\expandafter\ifx\csname natexlab\endcsname\relax\def\natexlab#1{#1}\fi

\bibitem[Ahuja \& Rowley(2010)]{Ahuja:JFM10}
{\sc Ahuja, S. \& Rowley, C.~W.} 2010 Feedback control of unstable steady
  states of flow past a flat plate using reduced-order estimators. {\em J.
  Fluid Mech.\/} {\bf 645}, 447--478.

\bibitem[Bagheri {\em et~al.\/}(2009)Bagheri, H{\oe}pffner, Schmid \&
  Henningson]{Bagheri:AMR09}
{\sc Bagheri, S., H{\oe}pffner, J., Schmid, P. \& Henningson, D.} 2009
  Input-output analysis and control design applied to a linear model of
  spatially developing flows. {\em Applied Mechanics Reviews\/} {\bf 62}~(2).

\bibitem[Batchelor(2000)]{batchelor2000introduction}
{\sc Batchelor, G.} 2000 {\em An introduction to fluid dynamics\/}. Cambridge
  university press.

\bibitem[Bencz\'{u}r \& Karger(1996)]{Benczur96}
{\sc Bencz\'{u}r, A. \& Karger, D.~R.} 1996 Approximating $s-t$ minimum cuts in
  $\mathcal{O}(n^2)$ time. In {\em Proceedings of the twenty-eighth annual ACM
  symposium on Theory of computing\/}, pp. 47--55. {ACM}.

\bibitem[Berkooz {\em et~al.\/}(1993)Berkooz, Holmes \&
  Lumley]{berkooz1993proper}
{\sc Berkooz, G., Holmes, P. \& Lumley, J.} 1993 The proper orthogonal
  decomposition in the analysis of turbulent flows. {\em Annual review of fluid
  mechanics\/} {\bf 25}~(1), 539--575.

\bibitem[Bollob{\'a}s(1998)]{Bollobas98}
{\sc Bollob{\'a}s, B.} 1998 {\em Modern graph theory\/}. Springer.

\bibitem[Cauchemeza {\em et~al.\/}(2011)Cauchemeza, A, Marchbanks, Fagan,
  S.~Ostroff, Swerdlow \& working group]{Cauchemeza:PNAS11}
{\sc Cauchemeza, S., A, Bhattarai, Marchbanks, T.~L., Fagan, R.~P., S.~Ostroff,
  N. M.~Ferguson, Swerdlow, D. \& working group, Pennsylvania~{H1N1}} 2011 Role
  of social networks in shaping disease transmission during a community
  outbreak of 2009 {H1N1} pandemic influenza. {\em Proceedings of the National
  Academy of Sciences\/} {\bf 108}~(7), 2825--2830.

\bibitem[Chen(2004)]{wai2004electrical}
{\sc Chen, W.~K.} 2004 {\em The Electrical Engineering Handbook\/}. Elsevier
  Science.

\bibitem[Chung(1997)]{Chung:1997}
{\sc Chung, F.} 1997 {\em Spectral Graph Theory\/}. American Mathematical
  Society.

\bibitem[Cottet \& Koumoutsakos(2000)]{Cottet00}
{\sc Cottet, G.-H. \& Koumoutsakos, P.~D.} 2000 {\em Vortex methods: theory and
  practice\/}. Cambridge Univ. Press.

\bibitem[Duarte-{C}avajalino {\em et~al.\/}(2012)Duarte-{C}avajalino,
  Jahanshad, Lenglet, Mc{M}ahon, de~{Z}ubicaray, Martin, Wright, Thompson \&
  Sapiro]{DuarteCavajalino:Neuroimage12}
{\sc Duarte-{C}avajalino, J.~M., Jahanshad, N., Lenglet, C., Mc{M}ahon, K.~L.,
  de~{Z}ubicaray, G.~.I., Martin, N.~G., Wright, M.~J., Thompson, P.~M. \&
  Sapiro, G.} 2012 Hierarchical topological network analysis of anatomical
  human brain connectivity and differences related to sex and kinship. {\em
  NeuroImage\/} {\bf 59}, 3784--3804.

\bibitem[Ellens {\em et~al.\/}(2011)Ellens, Spieksma, Van~Mieghem, Jamakovic \&
  Kooij]{ellens2011effective}
{\sc Ellens, W, Spieksma, FM, Van~Mieghem, P, Jamakovic, A \& Kooij, RE} 2011
  Effective graph resistance. {\em Linear algebra and its applications\/} {\bf
  435}~(10), 2491--2506.

\bibitem[Glass {\em et~al.\/}(2006)Glass, Glass, Beyeler \& Min]{Glass:EID06}
{\sc Glass, R.~J., Glass, L.~M., Beyeler, W.~E. \& Min, H.~J.} 2006 Targeted
  social distancing design for pandemic influenza. {\em Emerg. Infect.
  Diseases\/} {\bf 12}~(11), 1671--1681.

\bibitem[Greengard \& Rokhlin(1987)]{Greengard:JCP87}
{\sc Greengard, L. \& Rokhlin, V.} 1987 A fast algorithm for particle
  summations. {\em J. Comput. Phys.\/} {\bf 73}, 325--348.

\bibitem[Hemati {\em et~al.\/}(2014)Hemati, Eldredge \&
  Speyer]{hemati2014improving}
{\sc Hemati, M., Eldredge, J.~D. \& Speyer, J.~L.} 2014 Improving vortex models
  via optimal control theory. {\em Journal of Fluids and Structures\/} .

\bibitem[Holmes {\em et~al.\/}(1996)Holmes, Lumley \& Berkooz]{Holmes96}
{\sc Holmes, P., Lumley, J.~L. \& Berkooz, G.} 1996 {\em Turbulence, coherent
  structures, dynamical systems and symmetry\/}. Cambridge Univ. Press.

\bibitem[Kaiser {\em et~al.\/}(2014)Kaiser, Noack, Cordier, Spohn, Segond,
  Abel, Daviller \& Niven]{kaiser2013cluster}
{\sc Kaiser, E., Noack, B.~R., Cordier, L., Spohn, A., Segond, M., Abel, M.,
  Daviller, G. \& Niven, R.~K.} 2014 Cluster-based reduced-order modelling of a
  mixing layer. {\em Journal of Fluid Mechanics\/} {\bf 754}, 365--414.

\bibitem[Kelner \& Levin(2011)]{kelner2011spectral}
{\sc Kelner, J. \& Levin, A.} 2011 Spectral sparsification in the
  semi-streaming setting. {\em Leibniz International Proceedings in Informatics
  (LIPIcs) series\/} {\bf 9}, 440--451.

\bibitem[Klein \& Randi{\'c}(1993)]{klein1993resistance}
{\sc Klein, Douglas~J \& Randi{\'c}, M} 1993 Resistance distance. {\em Journal
  of Mathematical Chemistry\/} {\bf 12}~(1), 81--95.

\bibitem[Leonard(1980)]{leonard1980vortex}
{\sc Leonard, Anthony} 1980 Vortex methods for flow simulation. {\em Journal of
  Computational Physics\/} {\bf 37}~(3), 289--335.

\bibitem[Lloyd-{S}mith {\em et~al.\/}(2005)Lloyd-{S}mith, Schreiber, Kopp \&
  Getz]{LloydSmith:Nature05}
{\sc Lloyd-{S}mith, J., Schreiber, S., Kopp, P. \& Getz, W.} 2005
  Superspreading and the effect of individual variation on disease emergence.
  {\em Nature\/} {\bf 438}, 355--359.

\bibitem[Ma {\em et~al.\/}(2009)Ma, Ahuja \& Rowley]{Ma:TCFD09}
{\sc Ma, Z., Ahuja, S. \& Rowley, C.~W.} 2009 Reduced order models for control
  of fluids using the eigensystem realization algorithm. {\em Theo. Comp. Fluid
  Dyn.\/} {\bf 25}~(1), 233--247.

\bibitem[Mieghem(2011)]{van2011graph}
{\sc Mieghem, P.~V.} 2011 {\em Graph spectra for complex networks\/}. Cambridge
  University Press.

\bibitem[Mohar(1991)]{mohar1991laplacian}
{\sc Mohar, B.} 1991 The {Laplacian} spectrum of graphs. In {\em Graph theory,
  combinatorics, and applications\/} (ed. Y.~Alavi, G.~Chartrand, O.~Ollermann
  \& A.~Schwenk), pp. 871--898. New York: Wiley.

\bibitem[Morris(1993)]{Morris:SMR93}
{\sc Morris, M.} 1993 Epidemiology and social networks - modeling structured
  diffusion. {\em Sociol. Method Res.\/} {\bf 22}, 99--126.

\bibitem[Newman(2004)]{newman2004fast}
{\sc Newman, M. E.~J.} 2004 Fast algorithm for detecting community structure in
  networks. {\em Physical review E\/} {\bf 69}, 066133.

\bibitem[Newman(2010)]{Newman10}
{\sc Newman, M. E.~J.} 2010 {\em Networks: an introduction\/}. Oxford Univ.
  Press.

\bibitem[Newton(2001)]{Newton01}
{\sc Newton, P.~K.} 2001 {\em The $N$-vortex problem: analytical techniques\/},
  {\em Applied Mathematical Sciences\/}, vol. 145. Springer.

\bibitem[Noack {\em et~al.\/}(2005)Noack, Papas \& Monkewitz]{Noack:JFM05}
{\sc Noack, B.~R., Papas, P. \& Monkewitz, P.~A.} 2005 The need for a
  pressure-term representation in empirical {Galerkin} models of incompressible
  shear flows. {\em J. Fluid Mech.\/} {\bf 523}, 339--365.

\bibitem[Owen {\em et~al.\/}(2013)Owen, Li, Ziv, Strominger, Gold, Bukhpun,
  Wakahiro, Friedman, Sherr \& Mukherjee]{Owen:Neuroimage13}
{\sc Owen, J.~P., Li, Y.-O., Ziv, E., Strominger, Z., Gold, J., Bukhpun, P.,
  Wakahiro, M., Friedman, E.~J., Sherr, E.~H. \& Mukherjee, P.} 2013 The
  structural connectome of the human brain in agenesis of the corpus callosum.
  {\em NeuroImage\/} {\bf 70}, 340--355.

\bibitem[Peleg \& Ullman(1989)]{peleg1989optimal}
{\sc Peleg, D. \& Ullman, J.} 1989 An optimal synchronizer for the hypercube.
  {\em SIAM Journal on computing\/} {\bf 18}~(4), 740--747.

\bibitem[Porter {\em et~al.\/}(2005)Porter, Mucha, Newman \&
  Warmbrand]{Porter:PNAS05}
{\sc Porter, M.~A., Mucha, P.~J., Newman, M. E.~J. \& Warmbrand, C.~M.} 2005 A
  network analysis of committees in the {U.S.} {House of Representatives}. {\em
  Proc. Nat. Acad. Sci.\/} {\bf 102}~(20), 7057--7062.

\bibitem[Robinson {\em et~al.\/}(2012)Robinson, Cohen \&
  Colijn]{Robinson:TPB12}
{\sc Robinson, K., Cohen, T. \& Colijn, C.} 2012 The dynamics of sexual contact
  networks: effects on disease spread and control. {\em Theo. Popul. Bio.\/}
  {\bf 81}, 89--96.

\bibitem[Rowley {\em et~al.\/}(2004)Rowley, Colonius \&
  Murray]{Rowley2004model}
{\sc Rowley, C., Colonius, T. \& Murray, R.} 2004 Model reduction for
  compressible flows using {POD} and {G}alerkin projection. {\em Physica D:
  Nonlinear Phenomena\/} {\bf 189}~(1), 115--129.

\bibitem[Rowley(2005)]{Rowley:IJBC05}
{\sc Rowley, C.~W.} 2005 Model reduction for fluids, using balanced proper
  orthogonal decomposition. {\em Int. J. Bif. Chaos\/} {\bf 15}~(3), 997--1013.

\bibitem[Saffman(1992)]{Saffman92}
{\sc Saffman, P.~G.} 1992 {\em Vortex dynamics\/}. Cambridge Univ. Press.

\bibitem[Salath{\'e} \& Jones(2010)]{Salathe:PLOSCB10}
{\sc Salath{\'e}, M. \& Jones, J.~H.} 2010 Dynamics and control of diseases in
  networks with community structure. {\em PLoS Comput. Bio.\/} {\bf 6}~(4),
  e1000736.

\bibitem[Spielman \& Srivastava(2011)]{Spielman:SIAMJC11b}
{\sc Spielman, D.~A. \& Srivastava, N.} 2011 Graph sparsification by effective
  resistances. {\em SIAM J. Comput.\/} {\bf 40}~(6), 1913--1926.

\bibitem[Spielman \& Teng(2011)]{Spielman:SIAMJC11a}
{\sc Spielman, D.~A. \& Teng, S.-H.} 2011 Spectral sparsification of graphs.
  {\em SIAM J. Comput.\/} {\bf 40}~(4), 981--1025.

\bibitem[Srivastava(2010)]{srivastava2010spectral}
{\sc Srivastava, N.} 2010 Spectral sparsification and restricted invertibility.
  PhD thesis, Yale University.

\bibitem[Wang \& Eldredge(2013)]{wang2013low}
{\sc Wang, C \& Eldredge, J} 2013 Low-order phenomenological modeling of
  leading-edge vortex formation. {\em Theoretical and Computational Fluid
  Dynamics\/} {\bf 27}~(5), 577--598.

\end{thebibliography}

\end{document}